\documentclass[aps,prd,10pt,nofootinbib,twocolumn,eqsecnum,showpacs,showkeys,superscriptaddress,preprintnumbers,altaffilletter]{revtex4-2}
\usepackage{graphicx}
\usepackage{hyperref}
\usepackage{subcaption}
\usepackage{amsmath}
\usepackage{multirow}
\usepackage{makecell}
\usepackage{dcolumn}
\usepackage{amssymb}
\usepackage{amsfonts}
\usepackage{amsbsy}
\usepackage{tabularx}

\begin{document}

\title{Forecasting constraints on cosmology and modified gravitational-wave
    propagation by combining strongly lensed gravitational waves and
    galaxy surveys}

\date{\today}

\author{Anson Chen}
\email{chena@ucas.ac.cn}  
\affiliation{International Center for Theoretical Physics Asia-Pacific, University of Chinese Academy of Sciences, 100190 Beijing, China}

\author{Jun Zhang}
\affiliation{International Center for Theoretical Physics Asia-Pacific, University of Chinese Academy of Sciences, 100190 Beijing, China}
\affiliation{Taiji Laboratory for Gravitational Wave Universe, University of Chinese Academy of Sciences, 100049 Beijing, China}

\begin{abstract}

Gravitational lensing of gravitational wave (GW) will become the next frontier in studying cosmology and gravity. While time-delay cosmography using quadruply lensed GW events associated with optical images of the lens systems can provide precise measurement of the Hubble constant ($H_0$), they are considered to be much rarer than doubly lensed events. In this work, we analyze time-delay cosmography with doubly lensed GW events for the first time. We generate mock doubly lensed GW events with designed sensitivity of the LIGO-Virgo-KAGRA (LVK) O5 network, with LIGO post-O5 upgrade, and with Einstein Telescope (ET) + Cosmic Explorer (CE) respectively, and select the events that can be associated with future galaxy surveys. Over 1000 realizations, we find an average of $\sim 0.2(2.3)$ qualified events with the LVK O5(post-O5) network. Whereas with the ET+CE network, we find an average of 80.9 qualified events over 100 realizations. Using the Singular Isothermal Sphere (SIS) lens model, we forecast the constraints on cosmological parameters and modified GW propagation by combining time-delay cosmography and the standard siren approach. The average posterior gives a constraint on $H_0$ with a relative uncertainty of $14\%$, $10\%$ and $0.42\%$ in the $\Lambda$CDM model for the LVK O5, LVK post-O5, and ET+CE network, respectively. While the LVK network gives uninformative constraints on the $(w_0,w_a)$ dynamical dark energy model, the ET+CE network yields a moderate constraint of $w_0=-1.02^{+0.31}_{-0.25}$ and $w_a=0.39^{+1.01}_{-1.55}$. In addition, our method can provide precise constraints on modified GW propagation effects jointly with $H_0$. 
\end{abstract}

\maketitle

\section{Introduction} 

Despite of the great success of the Lambda Cold Dark Matter ($\Lambda$CDM) standard cosmology model, it is facing many challenges today. For instance, the second data release of the Dark Energy Spectroscopic Instrument (DESI) indicates that a dynamical dark energy model is favored by the Baryon acoustic oscillation (BAO) data, combined with data from the cosmic microwave background (CMB) and supernovae observations \cite{DESI:2025zgx,DESI:2025wyn}. It challenges the assumption of the $\Lambda$CDM model, where dark energy is interpreted as the Cosmological Constant. In addition, the Hubble tension problem has long existed between the CMB data and the local universe measurements, e.g. $H_0=67.49\pm0.50~{\rm km~s}^{-1}{\rm Mpc}^{-1}$ from the Planck mission \cite{Planck:2018vyg} and $H_0=74.03\pm1.42~{\rm km~s}^{-1}{\rm Mpc}^{-1}$ from the SH0ES collaboration \cite{Riess:2019cxk}, which may hint for new physics beyond the $\Lambda$CDM model (see review in \cite{Abdalla:2022yfr}). To further examine whether such observational anomalies are caused by systematics or unknown fundamental physics, more data from various aspects of observations are required in the near future.

On the other hand, gravitational wave (GW) detection has provided a unique laboratory in cosmological research. The latest Gravitational-Wave Transient Catalog 4.0 (GWTC-4) released by the LIGO-Virgo-KAGRA collaboration (LVK) reported 128 new compact binary coalescence (CBC) events from the first part of the fourth LVK observing run (O4a) \cite{LIGOScientific:2025slb}, which adds up the total number of CBC events to 218. These events can provide an independent measurement of $H_0$, as well as cosmological propagation effects, with the standard siren method \cite{Schutz1986,LIGOScientific:2017adf,LIGOScientific:2021aug,Mastrogiovanni:2023emh,Gray:2023wgj,LIGOScientific:2025jau,Jin:2025dvf}. For instance, the combination of the GW170817 bright siren event and the dark siren analysis for 142 of the GWTC-4 events yields $H_0=76.6^{+13.0}_
{-9.5}~{\rm km~s}^{-1}{\rm Mpc}^{-1}$ \cite{LIGOScientific:2025jau}. As more CBC events will be detected by the next-generation (XG) detectors in the future, such as Einstein Telescope (ET) \cite{Branchesi:2023mws} and Cosmic Explorer (CE) \cite{Evans:2021gyd}, the error of $H_0$ measurement by the dark siren method can be reduced to lower than 1\% \cite{Chen:2024gdn,Chen:2025qsl}.
With such precision in $H_0$ measurements, GWs are expected to provide a new perspective in solving the Hubble tension problem in the near future.

Apart from the standard siren method, $H_0$ can also be measured with GW detections in other aspects, in particular the gravitational lensing of GWs \cite{Sereno:2011ty,Hannuksela:2020xor,Huang:2023prq,Poon:2024zxn,Chen:2025xeg}. The time-delay cosmography with strong lensing has been applied in electromagnetic (EM) observations in the past few decades \cite{Refsdal:1964blz,Treu:2016ljm,H0LiCOW:2019pvv,Birrer:2020tax,Birrer:2022chj,Suyu:2023jue}. For example, the TDCOSMO Collaboration obtained the latest constraints $H_0=71.6^{+3.9}_{-3.3}~{\rm km~s}^{-1}{\rm Mpc}^{-1}$ with 8 strongly lensed quasars \cite{TDCOSMO:2025dmr}. On the other hand, GW signals can also be lensed by foreground galaxies, and result in multiple images for the same event. It has been proposed that these strongly lensed GW events can provide accurate measurements of $H_0$, in association with strongly lensed images of their host galaxies. However, it requires precise estimation of the Fermat potential, which can either be obtained by strongly lensed EM counterparts of the events \cite{Liao:2017ioi}, or by more than two images of the lensed events \cite{Hannuksela:2020xor,Poon:2024zxn,Chen:2025xeg}. These special events are considered to be rare in future detections, where 70\% of the strongly lensed events are expected to have only two images \cite{Li:2018prc}. Therefore, the chances to obtain cosmological constraints with these special strongly lensed events by LVK detections are very small.

In this work, we propose an alternative method to measure cosmological parameters with doubly lensed GW events, associating with images of the strong lensing systems observed by large-scale galaxy surveys. The latest survey performed by the Euclid satellite is expected to discover about 75,000 strong lensing systems \cite{Euclid:2025}, and the Legacy Survey of Space and Time (LSST) to be carried out by the Vera C. Rubin Observatory is also expected to detect about 70,000 strong lensing systems \cite{Shajib:2024yft}. The database of these strongly lensed galaxies can provide a precise map to identify the host galaxies of strongly lensed GW events. Using the Singular Isothermal Sphere (SIS) lens model, one can infer the Einstein radius $\theta_E$ from the angles between the two images of the source galaxy and the lens galaxy in a strong-lensing system. In addition to $\theta_E$, with the redshifts of the lens and the source galaxy, we can estimate $H_0$ using time-delay cosmography with strongly lensed GW signals, as well as dark energy or cosmological modified gravity effects described by the $(\Xi_0,n)$ parameterization or the Horndeski-class parameterization \cite{Belgacem:2018lbp,LISACosmologyWorkingGroup:2019mwx,Lagos:2019kds,Mukherjee:2020mha,Finke:2021aom,Mancarella:2021ecn,Leyde:2022orh,Chen:2023wpj,Narola:2023viz,Lin:2024pkr}. Apart from this, the luminosity distance of the GW event and the redshift of the source galaxy can provide a standard siren measurement of the cosmological parameters jointly with time-delay cosmography. The main uncertainty of time-delay cosmography with the SIS model comes from the impact parameter $y$, which can be jointly estimated from the two images of lensed GW signals. However, there exists a caveat that degeneracies between lens parameters and cosmology could occur when not considering more complex lens models for possible missing images, which requires more careful consideration for precision cosmology.

The structure of this paper is listed as followed. Section \ref{sec:theory} introduces the theoretical framework of strong lensing with the SIS model, cosmology models, and modified GW propagation. Section \ref{sec:method} describes our method to infer constraints on cosmological parameters. Section \ref{sec:simulation} presents the mock data simulation we perform to obtain the cosmological constraints, and the inference results are shown in Section \ref{sec:results}.
Finally, we summarize our work in Section \ref{sec:conclusion}.

\section{Theoretical framework}
\label{sec:theory}

\subsection{Strong-lensing with the SIS model}
\label{sec:sis}

In this work, we use the SIS lens model to approximate galaxy lenses. It has been shown in literature that the eccentricity and detailed radial-mass distribution of the lens galaxy make a relatively small impact to lens statistic \cite{Maoz:1993ix,Kochanek:1995ap,Mitchell:2004gw,Sereno:2011ty}. However, cosmological implication could be biased when using inaccurate lens models. In particular, for heavily elliptical lens galaxies, cosmological implication would be biased when using the SIS model. Still, the accurate reconstruction of the elliptical lens model relies on more than two images, which are rare among all lensed events. Nonetheless, we use the SIS model to analyse doubly lensed events as an ideal condition in this work, in order to forecast the constraints on cosmology with realistic lensed event population. Detailed study on the biases by lens galaxy ellipticity would be needed in future works.

\begin{figure}[t]
    \centering
    \includegraphics[width=\linewidth]{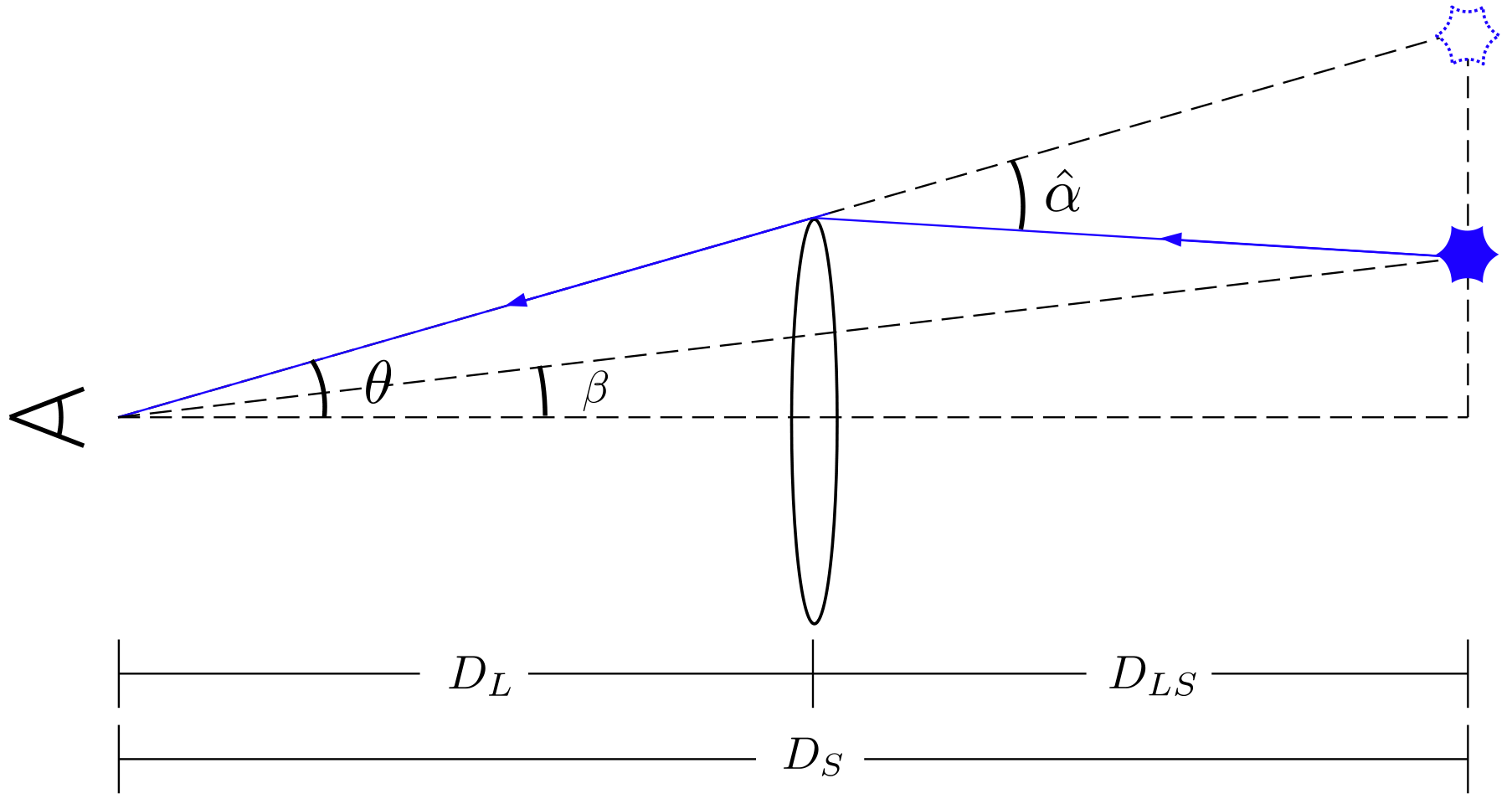}
    \caption{Schematic figure for strong-lensing with the thin-lens approximation. }
    \label{fig:lensing_sketch}
\end{figure}

The SIS model describes the galaxy mass distribution by assuming that the behavior of stars and other mass components of the galaxy are similar to ideal gas particles in an adiabatic system \cite{1987gady.book.....B}. The density profile of the SIS model is given by
\begin{equation}
    \rho(r) = \frac{\sigma_v^2}{2\pi G}\frac{1}{r^2},
\end{equation}
where $\sigma_v$ is the velocity dispersion. The Einstein radius of the SIS model is given by \cite{Takahashi_2003}
\begin{equation}
    \theta_E = 4\pi\frac{\sigma_v^2}{c^2}\frac{D_{\rm LS}}{D_{\rm S}},
\end{equation}
where $D_{\rm LS}$ and $D_{\rm S}$ are the angular diameter distances between the lens and the source, and between the observer and the source respectively. The lens mass for the SIS model is further given by \cite{Takahashi_2003}
\begin{equation}
    M_{L} = 4\pi\frac{\sigma_v^4}{Gc^2}\frac{D_{\rm L}D_{\rm LS}}{D_{\rm S}},
\end{equation}
where $D_{\rm L}$ is the angular diameter distance from the observer to the lens. Therefore one can obtain the relation 
\begin{equation}
\theta_E^2 = 4\pi \frac{GM_L}{c^2}\frac{D_{\rm LS}}{D_{\rm L}D_{\rm S}}.
\label{eq:theta_E_sis}
\end{equation}
In addition, a special feature of the SIS model is that the reduced deflection angle\footnote{The reduced deflection angle ${\alpha}$ relates to the deflection angle $\hat{\alpha}$ in Figure \ref{fig:lensing_sketch} by $\alpha=\hat{\alpha}D_{LS}/D_S$.} $\alpha$ equals to $\theta_E$, so that the lens equation becomes
\begin{equation}
    \theta_{\pm} = \beta \pm \theta_E,
    \label{eq:sis_lens_eq}
\end{equation}
where $\theta_\pm$ stands for the angles between the two images and the lens on the line-of-sight, and $\beta$ is the true source angular position (see Figure \ref{fig:lensing_sketch}). If $\beta<\theta_E$, there exists two images. Otherwise, there is only one image. Therefore, one can directly obtain the Einstein radius $\theta_E=(\theta_+ - \theta_-)/2$ for the doubly lensed system, which is half of the angle between the two images in observations. 

On the other hand, the time delay of the lensed signals is given by
\begin{equation}
    t_d(\vec\theta,\vec\beta) = \frac{1+z_L}{c}\frac{D_{\rm L}D_{\rm S}}{D_{\rm LS}}\left[\frac{1}{2}(\vec\theta-\vec\beta)^2-\psi(\vec\theta) \right],
\end{equation}
where $\psi(\vec\theta)$ is the effective potential of the lens, and it causes the Shapiro time delay of the signals. The time delay can also be written in terms of dimensionless quantities:
\begin{equation}
    t_d(\vec x,\vec y) = \frac{1+z_L}{c}\frac{D_{\rm L}D_{\rm S}}{D_{\rm LS}}\theta_E^2 T(\vec x,\vec y),
    \label{eq:time_delay}
\end{equation}
where $\vec x=\vec\theta/\theta_E$ and $\vec y=\vec\beta/\theta_E$ are dimensionless relative position of the image and the source. In the SIS system, the dimensionless time delay for the two images is given by \cite{Takahashi:2016jom}
\begin{equation}
    T_\pm = \mp y - \frac{1}{2}.
    \label{eq:sis_time_delay}
\end{equation}
The uncertainty of time delay measurement in EM strong lensing is a significant contribution to the uncertainty in cosmological constraint \cite{H0LiCOW:2019pvv}. Meanwhile, the arrival time of transient signals such as Gamma-ray bursts and GWs can be precisely measured, so the observed time delay measurement uncertainty in GW lensing is greatly reduced. 

For gravitationally lensed GWs, the waveform is modified by $\tilde{h}_{\rm L}(f)=F(\omega)\tilde{h}(f)$ in the frequency domain, where $F(\omega)$ is called the amplification factor ($\omega=2\pi f$). In the region of CBC GWs lensed by galaxies, the wavelength of GWs is much lower than the lens size and satisfies $|\Delta t_d \cdot \omega|\gg1$, so we can use the geometric optic approximation of $F(\omega)$, given by \cite{Takahashi_2003,Ezquiaga:2020gdt}
\begin{equation}
    F(\omega) \simeq \sum_j \sqrt{|\mu_j|} \exp\left(i\omega t_{d,j}-i~{\rm sign}(\omega)n_j\frac{\pi}{2} \right),
\end{equation}
where $\mu_j$ is the magnification factor for the $j$-th image, and $n_j=0,1,2$ is the Morse phase factor for the type I, II, III image respectively. In the SIS model, the magnification factor is given by \cite{Takahashi:2016jom}
\begin{equation}
    \mu_\pm = 1 \pm \frac{1}{y}.
    \label{eq:sis_magnification}
\end{equation}
For lensed GWs, the waveform amplitude is multiplied by $|\mu_\pm|$, which introduces bias in the observed GW luminosity distance $d_L^{\rm GW}$. Therefore, one can first obtain $y$ by the ratio of the lensed signal, 
\begin{equation}
    y=\frac{\mu_+/\mu_-+1}{\mu_+/\mu_--1}.
\end{equation}
so that the lensing magnification can be removed from the lensed signals with equation (\ref{eq:sis_magnification}), and the true $d_L^{\rm GW}$ can be obtained. In our analysis, we will jointly estimate $d_L^{\rm GW}$ and $y$ with the two images of lensed GW signals (as shown in Section \ref{sec:waveform_PE}). 

For higher $y$ within $y<1$, the amplitude of the second image of lensed GWs is suppressed, so its signal-to-noise ratio (SNR) could be below the detection threshold, typically $\rho_{\rm th}=8$. Fortunately, with sub-threshold signal searching techniques, lensed signals can be digged out with SNR down to 6, and down to 4 in rare cases \cite{McIsaac:2019use,Li:2019osa,LIGOScientific:2023bwz,Li:2023zdl,LIGOScientific:2025cwb}, which greatly improve the chance to find strongly-lensed signal pairs. Therefore, in our work, sub-threshold images of lensed GWs with network SNR larger than 6 are included in our analysis.

\subsection{Cosmology model and Modified Gravitational-wave propagation}
\label{sec:cosmology}

With strongly-lensed GWs, we can estimate cosmological parameters and modified gravity parameters that affect GW propagation. In a flat universe described by the Friedman-Lemaître-Robertson-Walker (FLRW) Metric, the luminosity distance of an EM source at redshift $z$ is given by \cite{Dodelson:2003ft}
\begin{equation}
    d_L = (1+z)\int_0^z \frac{c~{\rm d}z'}{H_0 E(z')}.
    \label{eq:luminosity_distance}
\end{equation}
For the $\Lambda$CDM model, $E(z)=\sqrt{\Omega_{m,0}(1+z)^3+\Omega_{\Lambda,0}}$, where $\Omega_{m,0}$ and $\Omega_{\Lambda,0}$ are the energy density of matter and dark energy today (the radiation energy density is negligible in the late-time universe). The angular diameter distance is related to the luminosity distance by $d_A=d_L/(1+z)^2$.

However, dark energy can be dynamical over time, instead of a constant in the $\Lambda$CDM model. A simple parameterization of the equation of state of dynamical dark energy is the Chevallier-Polarski-Linder (CPL) parametrization \cite{Chevallier:2000qy,Linder:2002et}, which depends on the scale factor $a$ by 
\begin{equation}
    w(a) = w_0+w_a(1+a).
\end{equation}
The corresponding $E(z)$ is given by
\begin{equation}
    E(z) = \sqrt{\Omega_{m,0}(1+z)^3 + \Omega_{\Lambda,0}  (1+z)^{3(1+w_0+w_a)} e^{-\frac{3w_az}{(1+z)}}}.
    \label{eq:Ez_w0wa}
\end{equation}
The $\Lambda$CDM model is recovered by $w_0=-1$ and $w_a=0$. The first two BAO data releases by DESI show a preference for the CPL dark energy model over the $\Lambda$CDM model by over 2.8$\sigma$ for the first time, combining with CMB data and supernovae data \cite{DESI:2025zgx,DESI:2025wyn}.

The nature of dark energy remains a mystery nowadays, yet many explanations to it have been proposed. Modified gravity is one of the popular dark energy theories. In a general modified gravity framework, the propagation of GW would deviate from General Relativity (GR), making GW a unique probe to dark energy. Meanwhile, the GW propagation speed has been tightly constrained by the multi-messenger event GW170817 to be consistent to GR\footnote{Some theories predict a transition in GW speed at a lower energy scale than the LVK detection band, but it can be constrained by LISA in the future \cite{deRham:2018red,LISACosmologyWorkingGroup:2022wjo,Baker:2022eiz}.} \cite{LIGOScientific:2017vwq,LIGOScientific:2017zic,LIGOScientific:2017ync}. Therefore, the remaining propagation effect without strong constraints is the damping effect\footnote{Though see \cite{Romano:2023ozy} for correlation between GW speed and damping effect.}, as described in the GW propagation equation \cite{Nishizawa:2017nef,Arai:2017hxj,Amendola:2017ovw,Belgacem:2018lbp,LISACosmologyWorkingGroup:2019mwx}:
\begin{equation}
    \tilde{h}_A'' + 2{\cal H}[1-\delta(\eta)]\tilde{h}_A' + c^2k^2\tilde{h}_A = 0,
    \label{eq:modified_propagation}
\end{equation}
where $\tilde{h}_A(\eta, k)$ is GW in Fourier space, the subscript $A$ denotes the + or $\times$ GW polarization, $\eta$ is the conformal time, $'$ denotes the derivative with respect to $\eta$, and ${\cal H} = a'/a$ is the conformal Hubble parameter. The term $\delta(\eta)$ is the extra damping effect induced by modified gravity, and GR is recovered at $\delta(\eta)=0$. By solving equation (\ref{eq:modified_propagation}), one can find the ratio between the modified GW luminosity distance and the EM luminosity distance as defined in equation (\ref{eq:luminosity_distance}),
\begin{equation}
    \frac{d_L^{\rm GW}(z)}{d_L^{\rm EM}(z)} = \exp\left[-\int_0^z \frac{{\rm d}z'}{1+z'}\delta(z') \right].
\end{equation}
The explicit form of $\delta(z)$ can be derived in specific modified gravity frameworks\footnote{See \cite{Zhu:2023rrx,Zhang:2024rel} for a more general GW damping term that includes frequency-dependent modification.}. But one widely-used general parameterization originated from non-local modified gravity models is given by \cite{Belgacem:2017cqo,Belgacem:2017ihm}
\begin{equation}
    \frac{d_L^{\rm GW}(z)}{d_L^{\rm EM}(z)} = \Xi_0+\frac{1-\Xi_0}{(1+z)^n},
    \label{eq:dGW_ratio_Xi0n}
\end{equation}
where $\Xi_0$ and $n$ are constants. At $z=0$, the distance ratio is 1, and it approaches the constant $\Xi_0$ as $z$ increases, which describes the deviation from GR consistent with dark-energy driven expansion of the universe at low redshift. This parameterization can be mapped into different modified gravity models (see \cite{LISACosmologyWorkingGroup:2019mwx} for some examples). A first constraint on $\Xi_0$ has been given by the dark siren method with GWTC-3 and GWTC-4 data \cite{Finke:2021aom,Leyde:2022orh,Chen:2023wpj,LIGOScientific:2025jau}, though the uncertainty is large at this stage.

Another widely tested model is the alpha-basis parameterization built upon the Horndeski scalar-tensor theory framework, where deviation from GR at the linear perturbation level can be described by 4 parameters: $\alpha_M,\alpha_K$, $\alpha_B$ and $\alpha_T$ \cite{Bellini:2014fua,Linder:2015rcz}. While $\alpha_T$ is the deviation of GW speed from $c$, $\alpha_M$ describes the running rate of the effective Planck mass, which leads to the extra damping factor by $\delta(z) = -\alpha_M(z)/2$ \cite{LISACosmologyWorkingGroup:2019mwx}. The evolution of $\alpha_M$ is often parameterized to be proportional to the growth of dark energy density, $\alpha_M(z)=c_M\Omega_\Lambda(z)/\Omega_{\Lambda,0}$. Taking the dark energy density evolution in the $\Lambda$CDM model, the distance ratio is given by
\begin{equation}
    \frac{d_L^{\rm GW}(z)}{d_L^{\rm EM}(z)} = \exp\left\{ \frac{c_M}{2\Omega_{\Lambda,0}}\ln\frac{1+z}{[\Omega_{m,0}(1+z)^3+\Omega_{\Lambda,0}]^{1/3}} \right\}.
    \label{eq:dGW_ratio_cM}
\end{equation}
The constant $c_M$ has been constrained by the CMB and large-scale structures \cite{Bellini:2015xja,Noller:2018wyv,Ishak:2024jhs}, as well as the dark siren method \cite{Ezquiaga:2021ayr,Leyde:2022orh,Chen:2023wpj,LIGOScientific:2025jau}. Constraint of $\Xi_0$ and $c_M$ parameterization has also been proposed using quadruply lensed GW events \cite{Finke:2021znb,Narola:2023viz}. Meanwhile, it has been shown that the geometric optic time delay for modified GW propagation in scalar-tensor theories remains the same as in GR \cite{Garoffolo:2019mna,Tasinato:2021wol}.

\section{Method}
\label{sec:method}

In this work, we aim to obtain cosmological constraints with the doubly-lensed GW events and the lensed images of their host galaxies from strong lensing catalogues. Using the enhanced localization technique with the effective detector network consisting of ground-based GW detectors at different spacetime due to time delay of strongly lensed signals, the sky localization area of strongly lensed GWs can be tightly constrained to $\Delta\Omega\sim0.01~{\rm deg}^2$ with 3-4 images detected by 3 XG detectors \cite{Chen:2025xeg}, which greatly enhance our chances to find the corresponding strongly lensed host galaxies from galaxy surveys. On the other hand, the localization for LVK detectors has lower accuracy, which is expected to be $\Delta\Omega\sim2~{\rm deg}^2$ for quadruply lensed events \cite{Janquart:2021qov}, and $\Delta\Omega\gtrsim10~{\rm deg}^2$ for doubly lensed events with current LVK detectors \cite{Uronen:2024bth}. In the case where there are more than one strongly lensed galaxies within the localization area, one can compare the Bayes factor between each system with the likelihood and identify the true lensed systems, which have been studied in previous works \cite{Hannuksela:2020xor,Wempe:2022zlk}. Here we consider the optimistic case where true strongly lensed systems are identified for GW lensed events. Such systems should be within galaxy survey footprints, and the lensed image angular positions can be resolved by the EM telescopes (See Section \ref{sec:lens_select} for lensed event selection). Otherwise, lensed GW events with host galaxy EM images outside galaxy catalogues cannot be analyzed with our method.

Given the redshift of the source galaxy and the GW luminosity distance $d_L^{\rm GW}$, we can constrain cosmological parameters via the standard siren method. However, the first analysis of lensed GWs will result in biases in $d_L^{\rm GW}$ because of magnification. Therefore we need to re-analyze lensed GWs once the lensed signal pairs are confirmed. By joint re-analysis of the two images of lensed GWs, we can obtain the posterior samples of the true $d_L^{\rm GW}$ and the impact parameter $y$ of lensed events. See Section \ref{sec:waveform_PE} for details of waveform parameter re-analysis.

In addition, we can obtain cosmological constraints via time delay cosmography of strongly lensed GWs. Given the observed time delay between GW signal images, $\Delta t=t_{d,2}-t_{d,1}$, as well as redshifts of the lens galaxy and the source galaxy, the Einstein radius $\theta_E$, and the impact parameter $y$, we can estimate cosmological parameters with equation (\ref{eq:time_delay}) and (\ref{eq:sis_time_delay}) using the SIS model. For the resolved image of a strong-lensing system in galaxy surveys, the angular separation between the two images of the source can be observed. Then one can obtain $\theta_E$ of the SIS lensing system by $\theta_E=(\theta_+-\theta_-)/2=\Delta\theta/2$, given the lens equation (\ref{eq:sis_lens_eq}). Therefore, we construct the log-likelihood on cosmological parameters ${\lambda}$ with $\chi^2$-statistics over $N$ lensed events, given by
\begin{equation}
\log{\cal L}(\lambda) = -\frac{1}{2}\sum_{i=1}^N \chi^2(\lambda,\Delta t_i^{\rm obs},\Delta\theta_i,z_{L,i},z_{S,i},y_i),
\label{eq:likelihood}
\end{equation}
where
\begin{align}
\chi^2(\lambda,&\Delta t_i^{\rm obs},\Delta\theta_i,z_{L,i},z_{S,i},y_i) = \left[\frac{d_{L,i}^{\rm obs}-d_L(\lambda,z_{S,i})}{\delta d_{L,i}^{\rm obs}} \right]^2 \nonumber\\
& + \left[\frac{\Delta t_i^{\rm obs}-\Delta t(\lambda,\Delta\theta_i,z_{L,i},z_{S,i},y_i)}{\delta (\Delta t)} \right]^2.
\label{eq:chi_square}
\end{align}
The uncertainty in time delay reconstruction is dominated by $\delta y$ from GW re-analysis, which typically gives relative error of $10\%-20\%$. The uncertainty in other factors, namely $\Delta\theta,z_L,z_S$, is in order of at most a few percents from galaxy surveys\footnote{The redshift uncertainty from spectroscopic survey and photometric survey is expected to be in order of $0.1\%$ and $1\%$ respectively (e.g. \cite{Euclid:2024yrr}; \cite{2009arXiv0912.0201L}). Although photometric redshift of the source galaxy is affected by light bending due to gravitational lensing, follow-up spectroscopic survey could provide more accurate redshift measurement.}. Moreover, uncertainty in GW arrival time in milliseconds is negligible compared to time delay in days, and we assume a perfectly smooth lens model so that time delay from small features in galaxy lens model is neglected. Therefore we have $\delta (\Delta t)/\Delta t \simeq \delta y/y$ in the likelihood. Using flat priors on $\lambda$, the posterior for $\lambda$ can then be obtained by Bayesian inference with Markov Chain Monte Carlo (MCMC) sampling.

The aim of this work is to forecast cosmological constraints given observed lensed GW pairs. In practice, however, one must also consider the detectability of lensed GW pairs, and weight the likelihood of observables by selection effects. This should be analyzed explicitly in future works.

\section{Simulation}
\label{sec:simulation}

To forecast the cosmological constraints with doubly lensed GW events, we first simulate mock data for strongly lensed GW events based on theoretical models for GW source population, merger rate, and lensing rate. Next, we filter out the events that can be associated with strong-lensing systems based on the resolution, the sky coverage and the depth of galaxy surveys. We then perform joint parameter estimation for lensed GW signal pairs. Finally, we perform MCMC sampling to obtain constraints on cosmological parameters. Details of simulation in each step are clarified in the following sections.

\subsection{Mock GW events}

Our first step of simulation is generating mock GW data for the LVK O5 stage, the post-O5 stage, and the next-generation detectors ET+CE, using the Python tool \texttt{GWSim} \cite{Karathanasis:2022hrb}. We assume that the O5 detector network consists of LIGO Observatory at Livingston (L1) and Hanford (H1), Virgo (V1) and KAGRA (K1), with their designed sensitivity curves\footnote{L1 and H1: https://dcc.ligo.org/DocDB/0149/T1800042/005/\\AplusDesign.txt\\
V1: https://git.ligo.org/publications/detectors/obs-scenarios-2019/blob/master/Scripts/Figure1/data/avirgo\_O5high\_NEW.\\txt\\
K1: https://git.ligo.org/publications/detectors/obs-scenarios-2019/blob/master/CurvesForSimulation/kagra\_128Mpc.txt} \cite{KAGRA:2013rdx} that we implement in \texttt{GWSim}. For the post-O5 detector network, we assume that LIGO India (A1) will join detections with LIGO designed sensitivity, so that there are 5 detectors in total. We also assume that post-O5 upgrades will be installed in L1 and H1 with the A\# sensitivity\footnote{https://dcc.ligo.org/LIGO-T2300041-v1/public/\\Asharp\_strain.txt}. For the ET+CE network, we use the orientation and the location of ET with the triangular configuration implemented in \texttt{Bilby}\footnote{https://git.ligo.org/lscsoft/bilby/-/blob/\\d37609ccc9a878750c841b4b93da990e63acffb2/bilby/gw/\\detector/detectors/ET.interferometer}, with a sensitivity curve for the 10 km arm length design\footnote{https://apps.et-gw.eu/tds/ql/?c=16492/ET10kmcolumns.txt}. We then assume two CE locating at L1 and H1 sites, with a sensitivity curve for the 40 km arm length design\footnote{https://dcc.cosmicexplorer.org/CE-T2000017/public/\\cosmic\_explorer\_strain.txt}. For each detector, we assume a duty cycle of 75\% in an ideal operation, and it applies to the three ET detectors together as a whole.

We generate the mock binary black hole (BBH) merger events using the ``Powerlaw + Double Peak" black hole mass distribution model and the Madau-Dickinson redshift evolution model, which are used in GWTC-4 cosmology paper \cite{LIGOScientific:2025jau}. We only consider BBH here, since they compose of most of the GW events in LVK detections, while the event rates of binary neutron star and neutron-star-black-hole merger have large uncertainty. The description of these models and our injected values are shown in Appendix \ref{app:model}. The luminosity distances of GW events are then generated from redshift with the fiducial $\Lambda$CDM model with $H_0=67.8~{\rm km~s}^{-1}{\rm Mpc}^{-1}$ and $\Omega_{m,0}=0.308$ from Planck 2018 TT, TE, EE+lowE+lensing+BAO results \cite{Planck:2018vyg}. We assume 5 years of observations for each detector network configuration, and that GW events distribute uniformly in the sky map. We select mock events with SNR $>11$, and proceed to lensed event generation.

\subsection{Lensed event selection}
\label{sec:lens_select}

We follow \cite{Sereno:2011ty} to compute the optical depth for GW sources, which gives the probability for GW events being lensed and detected at different redshifts (see detailed derivation in Appendix \ref{app:optical_depth}). The optical depth $\tau$ is given by
\begin{equation}
    \tau = \frac{F_*}{30}[D_S(1+z_S)]^3 y_{\rm max}^2 \bigg\{ 1-\frac{1}{7}\frac{\Gamma[(8+\alpha)/\beta]}{\Gamma[(4+\alpha)/\beta]}\frac{\Delta t_*}{T_{\rm obs}} \bigg\},
    \label{eq:tau}
\end{equation}
where
\begin{equation}
    F_* = 16\pi^3 n_{*,0} \bigg(\frac{\sigma_{*,0}}{c} \bigg)^4 \frac{\Gamma[(4+\alpha)/\beta]}{\Gamma[\alpha/\beta]};
\end{equation}
\begin{equation}
    \Delta t_* = 32\pi^2 \bigg(\frac{\sigma_{*,0}}{c} \bigg)^4 \frac{D_S}{c}(1+z_S)y_{\rm max}.
\end{equation}
$D_S$ and $z_S$ are the angular diameter distance and redshift of the source, $\Gamma$ is the Euler gamma function, and $T_{\rm obs}$ is the observation time of GW detectors. We adopt the modified Schechter function parameters as $n_{*,0}=8.0\times10^{-3}~h^3{\rm Mpc}^{-3}$, $\sigma_{*,0}=144~{\rm km}~{\rm s}^{-1}$, $\alpha=2.49$, and $\beta=2.29$, following previous study in \cite{Sereno:2011ty}
(also consistent with recent studies, e.g. in \cite{Choi:2006qg,Geng:2021tiz,2025MNRAS.537..779F}).

The maximal impact parameter $y_{\rm max}$ depends on the network SNR detection threshold $\rho_{\rm th}$ and the network SNR of the unlensed GW signal $\rho_{\rm UL}$ by 
\begin{equation}
    y_{\rm max} = \bigg[\bigg( \frac{\rho_{\rm th}}{\rho_{\rm UL}} \bigg)^2+1 \bigg]^{-1}
    \label{eq:y_max}
\end{equation}
In our simulation, we compute $y_{\rm max}$ with $\rho_{\rm UL}$ for each event and $\rho_{\rm th}=6$ as the network SNR cut for sub-threshold search. We then randomly select lensed events from the mock GW catalogues weighted by the optical depth with $y_{\rm max}$ for each event using equation (\ref{eq:tau}). This criterion of $y_{\rm max}$ indicates that our simulated lensed events always have $y<y_{\rm max}$ as their second images pass the detection threshold.

For the selected lensed events, we create the foreground galaxy lenses based on the stellar mass distribution of galaxies. We adopt the double-Schechter function to described the galaxy density profile in stellar mass:
\begin{equation}
    \Phi(M) = \exp \bigg(-\frac{M}{M^*} \bigg) \bigg[\Phi^*_1 \bigg(\frac{M}{M^*}\bigg)^{\alpha_1+1} +\Phi^*_2 \bigg(\frac{M}{M^*}\bigg)^{\alpha_2+1} \bigg].
    \label{eq:double_schechter}
\end{equation}
We adopt $\log (M^*/M_\odot)=10.79$, $\log (\Phi^*_1/h^3{\rm Mpc}^{-3})=-3.31$, $\log (\Phi^*_2/h^3{\rm Mpc}^{-3})=-2.01$, $\alpha_1=-1.69$, and $\alpha_2=-0.79$ from \cite{2016MNRAS.459.2150W}, and randomly draw lens galaxy stellar mass with probability weighted by equation (\ref{eq:double_schechter}). Since the ratio between baryonic matter and dark matter is 1:5 in the universe, and baryonic matter in galaxies includes stars and interstellar media, we adopt a representative value of $10\%$ for stellar mass contribution to galaxy total mass including dark matter halo. Hence we obtain lens galaxy total mass as 10 times of the drawn stellar mass. In addition, we draw lens redshifts $z_L$ with probability weighted by the differential comoving volume $dV_c/dz$, and draw impact parameter $y$ with probability weighted by $y$ in $(0,y_{\rm max})$.

Since our method relies on the resolved images of lensed host galaxies, we need to filter out the lensed events whose EM counterparts are observable and resolvable in galaxy surveys. We target the LSST survey, which will cover redshift up to 3 and a sky area of 18000 ${\rm deg}^2$ \cite{2009arXiv0912.0201L,LSSTDarkEnergyScience:2012kar}, which is about $43\%$ of the sky, so we only keep the same portion of events out of all mock events. In addition, the resolvable angle for the LSST camera is 0.2 arcsecond \cite{LSST_overview}. For the selected lensed events, we compute $\theta_E$ with equation (\ref{eq:theta_E_sis}) from $M_L$ and $D_L$, $D_S$, $D_{LS}$ converted from $z_L$ and $z_S$ with the fiducial $\Lambda$CDM model. Then we select resolvable lensed events with $\theta_\pm=(y\pm1)\theta_E>0.2''$. The observed time delay between images of each event is computed with lens model parameters using equation (\ref{eq:theta_E_sis}), (\ref{eq:time_delay}) and (\ref{eq:sis_time_delay}).

We simulate 1000 realizations for LVK O5 and post-O5 observations. A realization means that we generate random mock lensed events based on optical depth of GW events in our mock catalogue given a different random seed, and collect all events with resolvable EM images in LSST's footprint. For the O5 run, we obtain at least one lensed events in 331 realizations. Considering events in the LSST's footprint, the average lensed event number is 0.40 over all 1000 realizations. However, we only obtain resolvable lensed events by LSST in 152 realizations after filtering, with an average of 0.17 resolvable events across all realizations. For the post-O5 run, we obtain resolvable lensed events in 895 realizations. Compared to an average of 5.4 total lensed events within LSST coverage, there are only 2.3 resolvable events across all realizations. For the ET+CE network, we obtain an average of 191.8 total lensed events, and 80.9 resolvable events in all of the 100 realizations. In our simulation, the fraction of resolvable lensed events by LSST is 24\%-43\%, which is consistent to the lensed event localization study by \cite{Wempe:2022zlk}.

\subsection{Waveform parameter estimation}
\label{sec:waveform_PE}

For each resolvable lensed event detected by LVK, we perform waveform parameter estimation (PE) jointly for the two signal images. We assume that both signals from a lensed event have been analyzed individually prior to our analysis, so that parameters like sky location, spin, time and phase at merger have been measured, which can be fixed in the re-analysis. In reality, there could be large uncertainty in individual PE and degeneracy between parameters, so a more robust approach is to jointly re-analyze all parameters. However, to reduce computational cost in our analysis, we run joint PE on 5 parameters while fixing the rest as a proof of principle for our method, assuming that the true location is well recovered for individual images. We choose to jointly estimate chirp mass ${\cal M}_c$, mass ratio $q$, GW luminosity distance $d_L^{\rm GW}$, inclination angle $\iota$, and the impact parameter $y$. PE of ${\cal M}_c$ and $d_L^{\rm GW}$ from single images could be biased by the magnification factor due to lensing, so we need to find their true values from re-analyses with the SIS lens model. Meanwhile, we include mass ratio because it degenerates with ${\cal M}_c$, and inclination angle because it degenerates with $d_L^{\rm GW}$.

We first generate the unlensed waveform signal in frequency domain using the \texttt{IMRPhenomXPHM} waveform \cite{Pratten:2020ceb}, and then create the two lensed signal images magnified by $\mu_\pm(y)$ with equation (\ref{eq:sis_magnification}) and a Morse phase difference of $\pi/2$ between images. In reality, the Morse phase depends on the type of observed images, but we assume that the true Morse phase is found by phase difference between images from individual PE, and we use $\pi/2$ for all events for simplicity. For simplicity, we generate spin-free waveforms. The posterior of waveform parameters $\{\phi\}$ given image data $x_{j=1,2}$ in the joint analysis is thus given by
\begin{equation}
    P(\phi|x_1,x_2) = \int P(x_1,x_2|\phi) \Pi(\phi) {\rm d}\phi,
\end{equation}
where $\Pi(\phi)$ is the prior, and the joint log likelihood is given by
\begin{equation}
    \log P(x_1,x_2|\phi) = -\frac{1}{2} \sum_{j=1}^2 \langle \tilde{h}(x_j)-\tilde{h}(\phi)| \tilde{h}(x_j)-\tilde{h}(\phi) \rangle.
\end{equation}
Here $f$ is the frequency, $\tilde{h}(f)$ is the detector response to GW signals, given by 
\begin{equation}
    \tilde{h}(f)= \sum_k [F_+^k \tilde{h}_+(f) + F_\times^k \tilde{h}_\times(f)],
\end{equation}
where $F_+^k$ and $F_\times^k$ are the antenna pattern functions of the $k$ detector for the plus and cross GW polarization. The scalar product $\langle \tilde{h}(f)|\tilde{g}(f) \rangle$ is defined as 
\begin{equation}
    \langle \tilde{h}(f)|\tilde{g}(f) \rangle \equiv 2~{\rm Re}\bigg[\int {\rm d}f \frac{\tilde{h}(f)\tilde{g}^*(f)+\tilde{h}^*(f)\tilde{g}(f)}{{\rm Sn}(f)} \bigg],
\end{equation}
where ${\rm Sn}(f)$ is the power spectral density (PSD) of the detector noise. The priors on $d_L^{\rm GW}$ and $y$ are $\Pi(d_L^{\rm GW}) \propto (d_L^{\rm GW})^2$ and $\Pi(y)\propto y$ respectively, while the priors for other parameters are uniform. Finally, the posterior samples are inferred with \texttt{nessai} \cite{Williams:2021qyt,Williams:2023ppp}, which incorporates normalising flows in nested sampling within the \texttt{Bilby} package \cite{bilby_paper}.
\begin{figure}[t!]
    \centering
    \includegraphics[width=\linewidth]{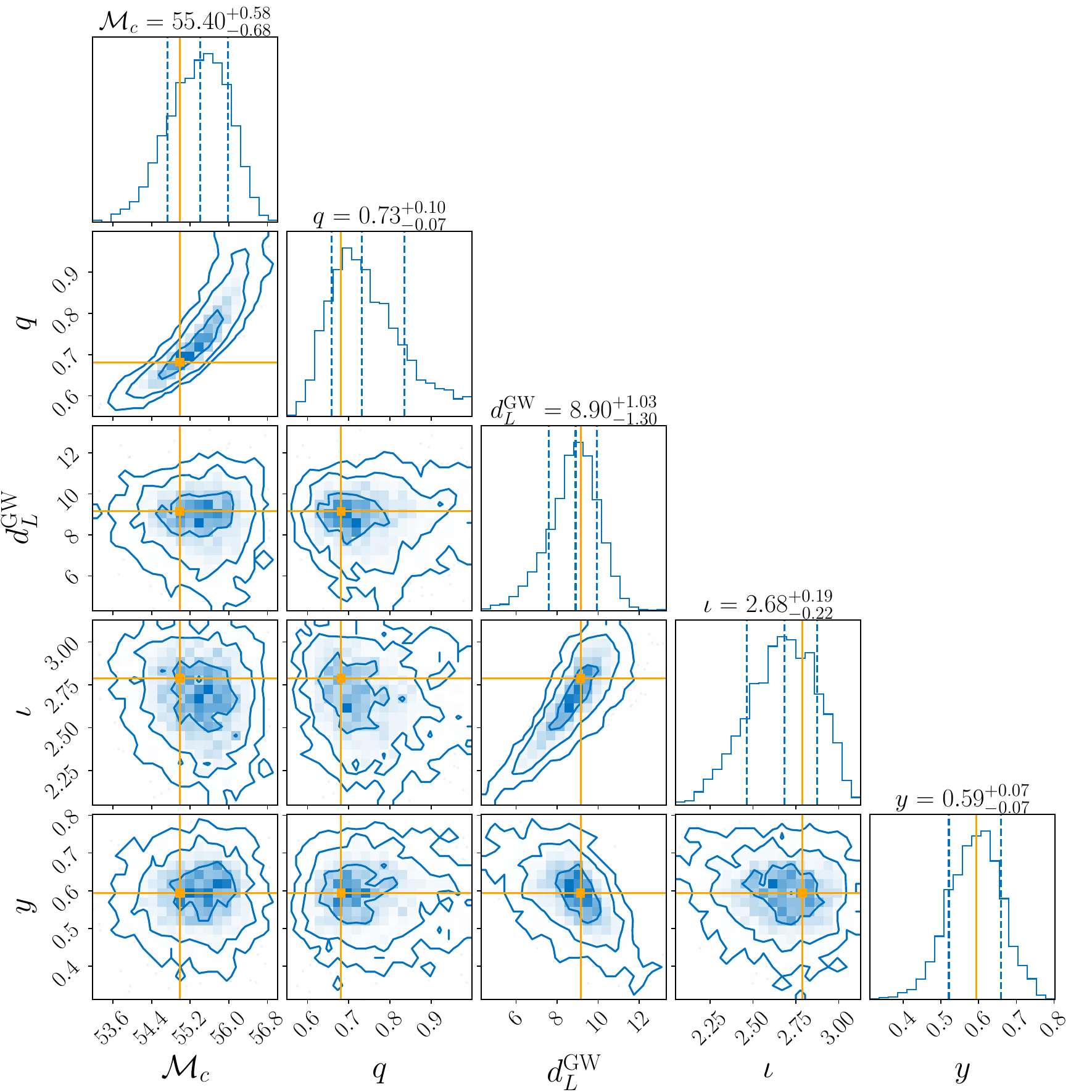}
    \caption{Posterior samples from joint re-analysis of one of the mock doubly lensed GW events detected by LVK O5 network using \texttt{nessai} nested sampling in \texttt{Bilby}. The estimated parameters include chirp mass ${\cal M}_c[M_\odot]$, mass ratio $q$, GW luminosity distance $d_L^{\rm GW}$[Gpc], inclination angle $\iota$, and impact parameter $y$. The orange lines show the injected values for parameters.}
    \label{fig:waveform_PE}
\end{figure}

Figure \ref{fig:waveform_PE} shows an example of the posterior samples from joint PE for lensed GW signals. We can see that, in spite of moderate degeneracy, the true $d_L^{\rm GW}$ and $y$ are well recovered, and their marginalized posteriors generally follow Gaussian distribution. We can therefore obtained the observables and their uncertainties from the mean and the standard deviation of the posterior samples, respectively. 

For the ET+CE network, however, the waveform parameter estimation is computationally expensive given a large event number. We thus
perform Fisher matrix analysis to forecast the uncertainty of the waveform parameters for the ET+CE events, which is given by the square root of the diagonal elements of the covariance matrix $\mathbf{\Sigma}^{ii}$ \cite{PhysRevD.46.5236,PhysRevD.49.2658}. The Fisher matrix is the inverse of the covariance matrix, constructed by $(\mathbf{\Sigma}^{ij})^{-1}=\langle \partial \tilde{h}/\partial\phi_i| \partial\tilde{h}/\partial\phi_j \rangle$. We numerically compute the partial derivatives of the waveform by finite differencing. The joint Fisher matrix for the two lensed GW signals is the sum of the Fisher matrices for the individual signals. Note that our ET+CE forecasts cannot be compared to our LVK forecasts directly, since we use Fisher matrix to approximate parameter errors for ET+CE instead of Bayesian inference.

\section{results}
\label{sec:results}

We infer the posterior for cosmological parameters with equation (\ref{eq:likelihood}) and (\ref{eq:chi_square}) from all qualified lensed GW events in each realization using MCMC sampling with the \texttt{emcee} package \cite{Foreman-Mackey:2012any}. For each realization, we estimate the posterior probability density function (PDF) from marginalized posterior samples with the Gaussian Kernel Density Estimation (KDE) method using \texttt{scipy.gaussian\_kde} under Scott’s Rule. Then we compute the average posterior PDF over all $N_{\rm real}$ realizations by\footnote{Realizations with no qualified lensed events are not included in averaging.}
\begin{equation}
    P_{\rm avg}(\lambda|x) = \frac{1}{N_{\rm real}} \sum_{i=1}^{N_{\rm real}} P_i(\lambda|x).
\end{equation}
In our analysis, we assume that we have identified the true lens system associated with the lensed GW events, and therefore use a Gaussian likelihood distribution for the lens parameter $y$. In reality, it would be more rigorous to marginalize over lens parameters given all possible lens systems within the localization area, which would allow for more weight to more confident GW-EM pairs. Such analysis will require tests with mock large-scale galaxy catalogue, which is beyond the scope of this paper.

We perform the inference in four different cosmology models, namely the $\Lambda$CDM model, the $w_0w_a$CDM model, the $(\Xi_0,n)$ modified gravity model, and the $\alpha$-basis scalar-tensor gravity model. The inference results are presented in the following subsections.

\subsection{$\Lambda$CDM model}

\begin{figure}[t]
    \centering
    \includegraphics[width=0.8\linewidth]{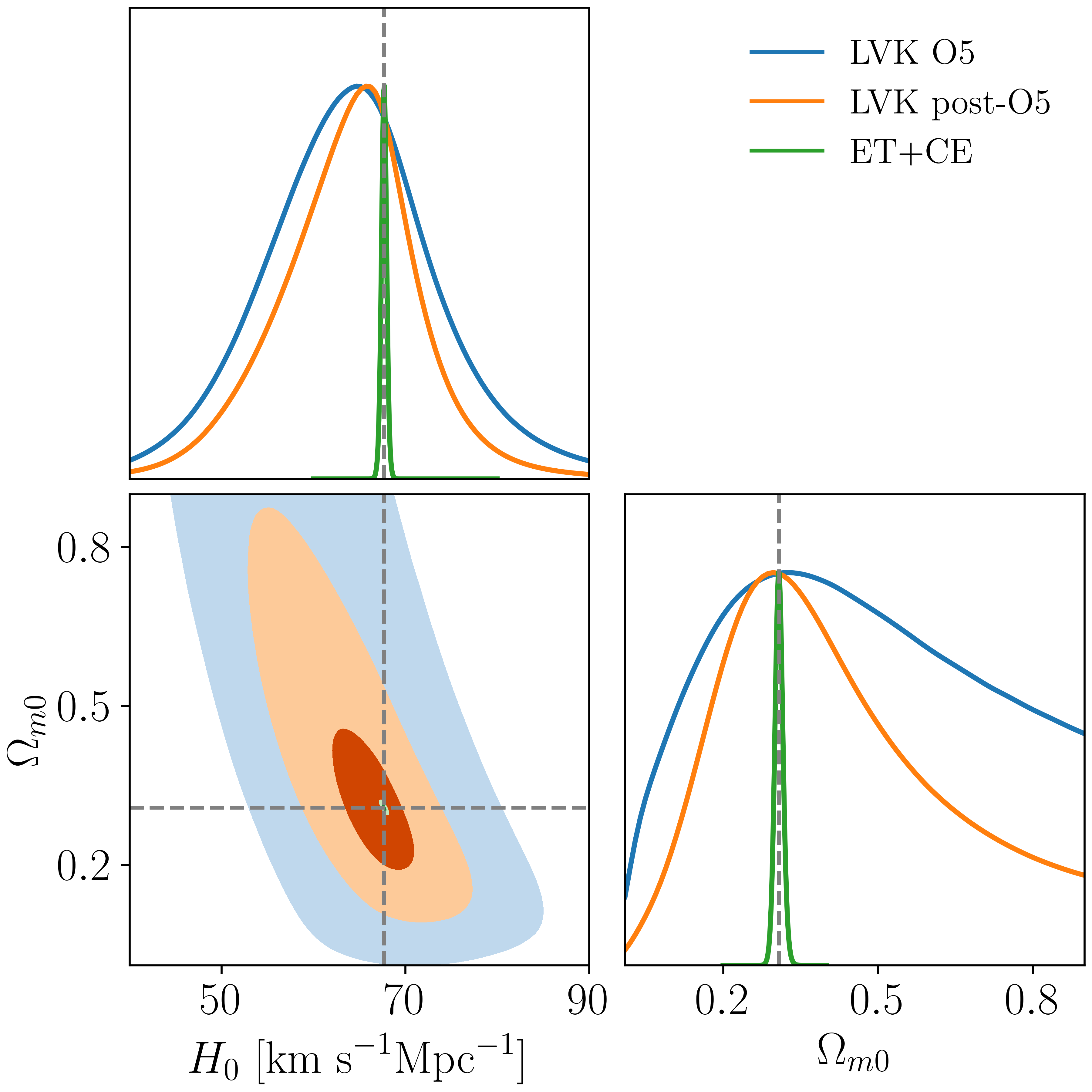}
    \caption{The figure shows the forecasted posterior on cosmological parameters in the $\Lambda$CDM model by jointly analyzing all lensed GW events in each realization, and averaging over all realizations, with different detector networks. The grey dashed lines indicate the parameter values in the fiducial $\Lambda$CDM model.}
    \label{fig:LCDM_corner}
\end{figure}

For the $\Lambda$CDM model, we estimate $H_0$ and $\Omega_{m,0}$, while assuming $\Omega_{\Lambda,0}=1-\Omega_{m,0}$ in the late universe. The averaged posteriors over all realizations are shown in Figure \ref{fig:LCDM_corner}. We see that for the LVK O5 network and the post-O5 network, the constraints on $H_0$ and $\Omega_{m,0}$ are weak, which is due to the limited number of lensed events. The relative 1$\sigma$ uncertainty of $H_0$ is around $14\%$ and $10\%$ for the O5 and the post-O5 network respectively (See exact number in Table \ref{tab:results}). For $\Omega_{m,0}$, the relative 1$\sigma$ uncertainty reaches $100\%$ and $73\%$ respectively. However, for the ET-CE network, the cosmological constraints are greatly improved, because the lensed event number is over 1 orders of magnitude larger than the LVK network, and the event SNRs are much larger. The relative 1$\sigma$ uncertainties on $H_0$ and $\Omega_{m,0}$ for the ET-CE network reduce to around $0.42\%$ and $2.6\%$ respectively. Such precision is comparable to Planck's measurement and other local-universe measurements, making doubly lensed GWs + strong lensing galaxy survey an accurate independent probe of $H_0$ that could provide a new insight to the Hubble tension. 

For comparison, the bright siren + dark siren constraints on $H_0$ with 142 of the GWTC-4 events yields an uncertainty of $\sim 15\%$ \cite{LIGOScientific:2025jau}, which can be achieved with 1-2 strongly lensed events with EM images. In addition, constraints on $H_0$ and $\Omega_{m,0}$ can also reach an uncertainty of $2.2\%$ and $5.5\%$ respectively by spectral siren + golden dark siren analysis with the ET+CE network in 1 year \cite{Chen:2024gdn}, and bright siren constraint on $H_0$ alone can reach an uncertainty below $1\%$ with $\sim 300$ events \cite{Chen:2024gdn}. Meanwhile, with a much smaller detection number, strongly lensed GW-EM pairs can provide a constraint at the same level as unlensed events. However, there exists different systematics in the two approaches. While the precision of the standard siren approach is affected by GW source population models and galaxy catalogue completeness, the strongly lensed GW approach relies on accurate lens signal identification, lens system localization, and lens model reconstruction. Therefore, these two approaches can complement each other in GW cosmology.

\subsection{$w_0w_a$CDM model}

\begin{figure}[t]
    \centering
    \includegraphics[width=\linewidth]{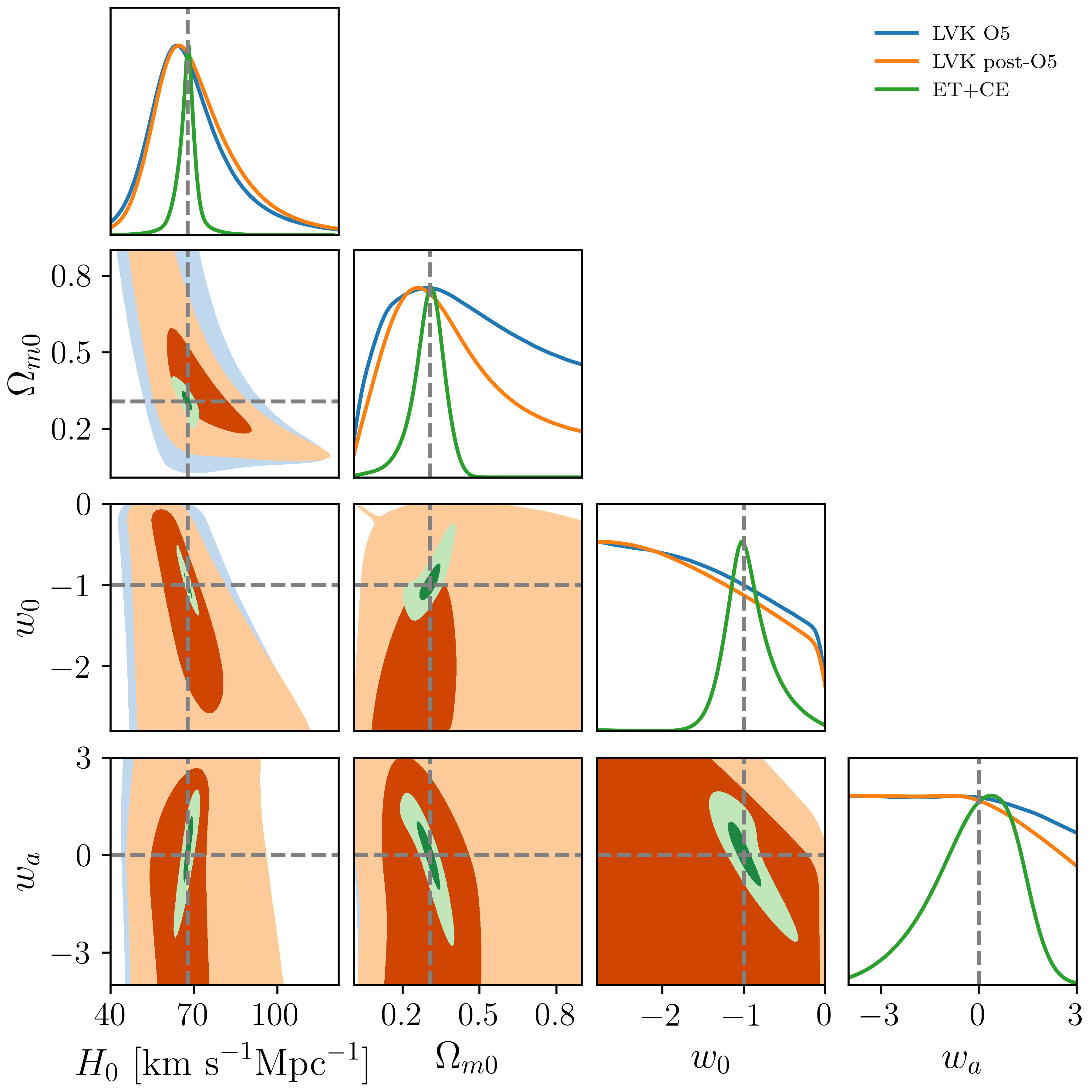}
    \caption{The figure shows the forecasted posterior on cosmological parameters in the $w_0w_a$CDM model by jointly analyzing all lensed GW events in each realization, and averaging over all realizations, with different detector networks. The grey dashed lines indicate the parameter values in the fiducial $\Lambda$CDM model.}
    \label{fig:w0waCDM_corner}
\end{figure}
In the dynamical dark energy model with the CPL parameterization, we estimate $H_0$, $\Omega_{m,0}$, $w_0$, and $w_a$, based on modification in $d_L^{\rm GW}$ with equation (\ref{eq:Ez_w0wa}). As shown by averaged posteriors over all realizations in Figure \ref{fig:w0waCDM_corner}, $w_0$ and $w_a$ are basically unconstrained with the LVK O5 and post-O5 network. On one hand, there are too few lensed events that can be associated with galaxy surveys for LVK sensitivity. On the other hand, the redshift range of LVK events are mostly at $z\lesssim2$, while precise constraints on $w_0$ and $w_a$ rely on high-redshift events. For the ET+CE network, we can see moderate constraints on $w_0$ and $w_a$, with uncertainties of $\sim 0.3$ for $w_0$, and around order of unity for $w_a$. It demonstrates the potential of doubly lensed GW events to constrain dynamical dark energy in the near future, though such constraint could be weaken if the true lenses deviate heavily from the SIS model. If the observation period of ET+CE can last longer, the uncertainty on $w_0$ and $w_a$ can be further reduced, making our method competitive to other EM probes.

\begin{figure}
    \centering
    \includegraphics[width=\linewidth]{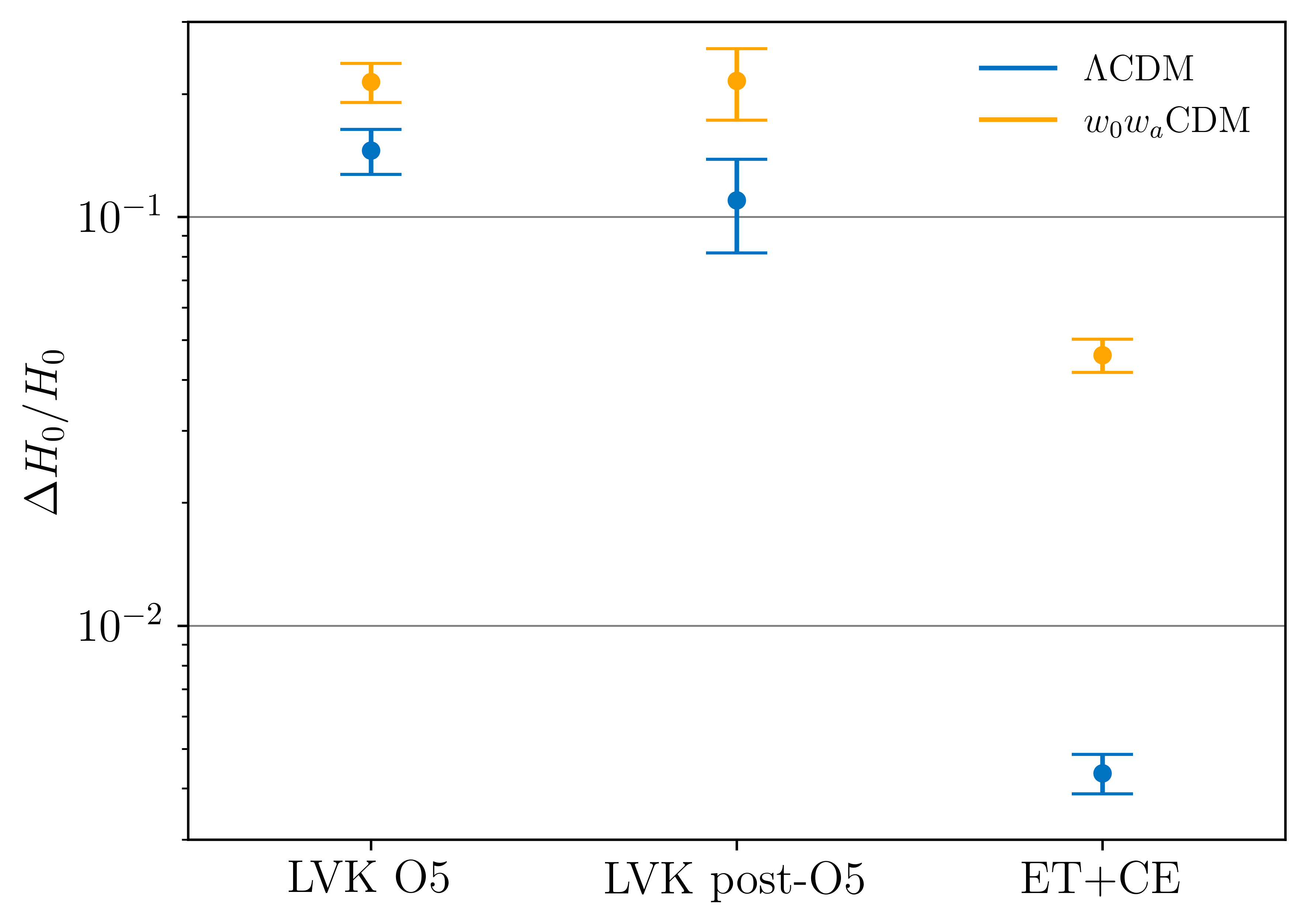}
    \caption{Fractional 1$\sigma$ uncertainty of $H_0$ in the $\Lambda$CDM model (blue) and $w_0w_a$CDM model (orange) with different detector networks. The upper and lower bars show the upper and lower $1\sigma$ uncertainty bounds respectively, and the circles represent their means.}
    \label{fig:H0_error}
\end{figure}

In addition, constraints on $H_0$ and $\Omega_{m,0}$ in the $w_0w_a$CDM model are also weaken compared to the $\Lambda$CDM model. The relative 1$\sigma$ uncertainty on $H_0$ becomes $21\%$ for both the LVK O5 and post-O5 network. Due to high degeneracy between $H_0$ and the dark energy parameters, as well as poor constraints on $w_0$ and $w_a$, the difference in $H_0$ constraints between the O5 and post-O5 network becomes insignificant compared to the decrease in accuracy. For the ET+CE network, the uncertainty on $H_0$ becomes $4.6\%$, which is increased by $\sim 10$ times. The evolution of the precision on $H_0$ measurement by different detector networks is shown in Figure \ref{fig:H0_error}. However, the uncertainty on $\Omega_{m,0}$ only increases slightly for the LVK O5 and post-O5 network, which may be due to the fact that $\Omega_{m,0}$ is already poor constrained by LVK. But it has climbed to $19\%$ for the ET+CE network, which is $\sim 7.4$ times the $\Omega_{m,0}$ uncertainty in the $\Lambda$CDM model.

\subsection{$(\Xi_0,n)$ modified gravity model}

\begin{figure}[t]
    \centering
    \includegraphics[width=\linewidth]{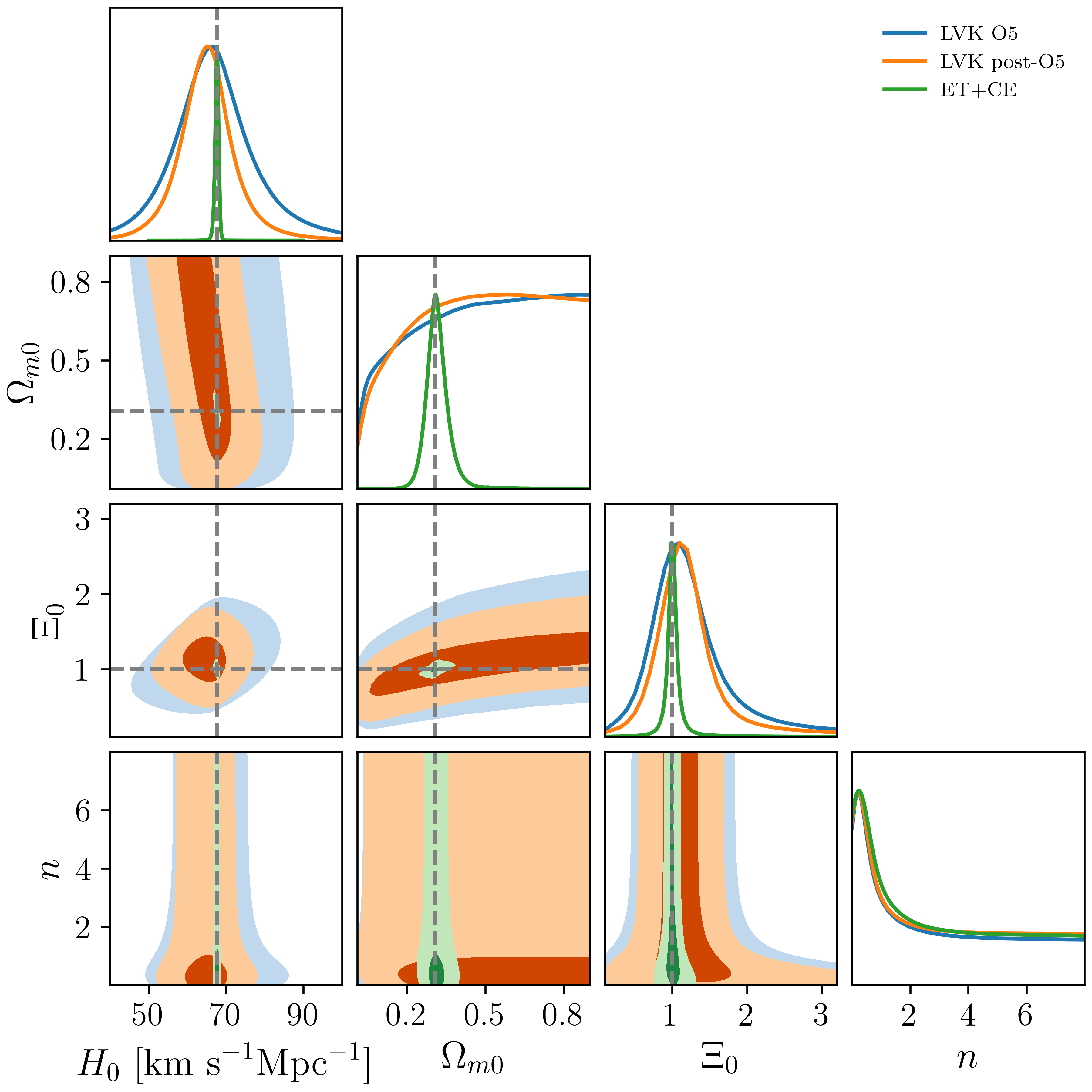}
    \caption{The figure shows the forecasted posterior on cosmological parameters in the $(\Xi_0,n)$ modified gravity model by jointly analyzing all lensed GW events in each realization, and averaging over all realizations, with different detector networks. The grey dashed lines indicate the parameter values in the fiducial $\Lambda$CDM model.}
    \label{fig:Xi0n_corner}
\end{figure}
We also analyze the constraints on modified gravity models with our method. For the $(\Xi_0,n)$ parameterization, we jointly estimate $H_0$, $\Omega_{m,0}$, $\Xi_0$ and $n$. The averaged posteriors over all realizations are presented in Figure \ref{fig:Xi0n_corner}. For the LVK network, we obtain moderate constraints on $H_0$ and $\Xi_0$, with a relative uncertainty in $H_0$ of $16\%$ and $11\%$ for the O5 network and the post-O5 network respectively, which is comparable to the $\Lambda$CDM case. But the relative uncertainty in $\Xi_0$ is $49\%$ and $36\%$ for the O5 network and the post-O5 network respectively. The constraint on $\Xi_0$ is not as tight as that on $H_0$, because $\Xi_0$ only modifies $d_L^{\rm GW}$ via equation (\ref{eq:dGW_ratio_Xi0n}) in the GW waveform, but not the angular diameter distances in lens systems from EM probes, while $H_0$ affects all these distances. However, we obtain poor constraints on $\Omega_{m,0}$ for the LVK network, probably due to its degeneracy with $H_0$ and $\Xi_0$. On the other hand, with the ET+CE network, we obtain strong constraints on $H_0$ and $\Xi_0$, and moderate constraint on $\Omega_{m,0}$. The relative uncertainties are $0.7\%$ and $8.0\%$ for $H_0$ and $\Xi_0$, and $12\%$ for $\Omega_{m,0}$. Finally, we find that the constraints on $n$ favor a value near the lower bound, but they are not converged at the higher bound, for all of the detector networks. It could be explained by the fact that the fiducial $\Lambda$CDM case corresponds to either $n=0$, or any value of $n$ for $\Xi_0=1$.

For comparison, the constraint on $\Xi_0$ has a relative uncertainty of $\sim 50\%$ from the GWTC-4 dark siren analysis \cite{LIGOScientific:2025jau}. Forecasts on the future dark siren constraint on $\Xi_0$ show an uncertainty of $\sim 10\%$ with LIGO/Virgo design sensitivity \cite{Mancarella:2021ecn}, and $\sim 0.4\%$ with the ET+CE network with precessing BBHs \cite{Lin:2024pkr}. Although the expected constraint for our method is slightly weaker than the dark siren constraint in the LVK O5 or the ET+CE era, the GW+EM strong-lensing constraint can still contribute to achieving a tighter combined constraint on modified gravity when used alongside other methods.

\subsection{$\alpha$-basis modified gravity model}

\begin{figure}[t]
    \centering
    \includegraphics[width=\linewidth]{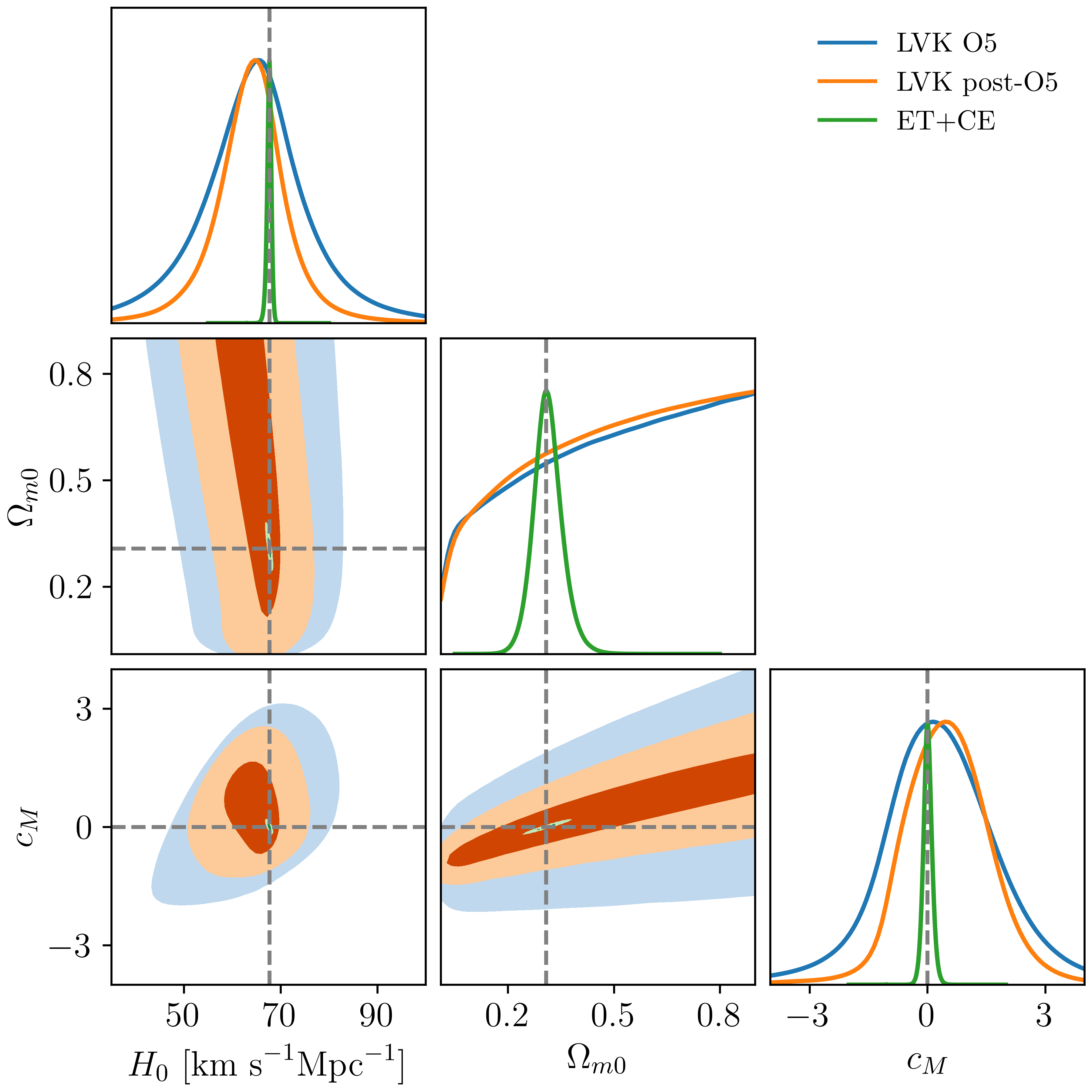}
    \caption{The figure shows the forecasted posterior on cosmological parameters in the scalar-tensor gravity model by jointly analyzing all lensed GW events in each realization, and averaging over all realizations, with different detector networks. The grey dashed lines indicate the parameter values in the fiducial $\Lambda$CDM model.}
    \label{fig:cM_corner}
\end{figure}
Apart from the general $(\Xi_0,n)$ modified gravity parametrization, we estimate constraints on the $\alpha$-basis parametrization as well. Here, $d_L^{\rm GW}$ is modified by equation (\ref{eq:dGW_ratio_cM}), so we jointly estimate $H_0$, $\Omega_{m,0}$ and $c_M$. As shown by the averaged posteriors over all realizations in Figure \ref{fig:cM_corner}, the constraints on $H_0$, $\Omega_{m,0}$ and $c_M$ are similar to those on $H_0$, $\Omega_{m,0}$ and $\Xi_0$ in the $(\Xi_0,n)$ parameterization. The relative uncertainty on $H_0$ is $11\%$, $7\%$, and $0.7\%$ for the LVK O5, post-O5, and ET+CE network, respectively. $\Omega_{m,0}$ is again unconstrained with the LVK network, but is moderately constrained with a relative error of $13\%$ for the ET+CE network. For $c_M$, the fiducial value in the $\Lambda$CDM model is $c_M=0$. The mean uncertainty of the upper bound and the lower bound of $c_M$ is $1.44$, $1.05$, and $0.12$, for the LVK O5, post-O5, and the ET+CE network, respectively. Similar to the degeneracy between $\Omega_{m,0}$ and $\Xi_0$ in the $(\Xi_0,n)$ model, we can also see degeneracy between $\Omega_{m,0}$ and $c_M$, but it is a bit stronger. This is likely because $d_L^{\rm GW}/d_L^{\rm EM}$ depends on both $\Omega_{m,0}$ and $c_M$ in the $\alpha$-basis model, while it is independent of $\Omega_{m,0}$ in the $(\Xi_0,n)$ model.

\renewcommand{\arraystretch}{2}
\begin{table*}[t]
    \centering
    \begin{tabular}{c|cc|cccc}
    \hline
    \hline
     & \multicolumn{2}{c}{$\Lambda$CDM} & \multicolumn{4}{c}{$w_0w_a$CDM} \\
    \hline
     & ~$H_0~[{\rm km}$ ${\rm s}^{-1}{\rm Mpc}^{-1}]$~ & $\Omega_{m,0}$ & ~$H_0~[{\rm km}$ ${\rm s}^{-1}{\rm Mpc}^{-1}]$~ & $\Omega_{m,0}$ & $w_0$ & $w_a$  \\
    \hline
    LVK O5 & $64.77^{+8.24}_{-10.60}$ & $0.325^{+0.382}_{-0.196}$~ & $63.83^{+15.17}_{-12.16}$ & $0.301^{+0.373}_{-0.211}$ & - & - \\
    \hline
    ~LVK post-O5~ & $65.88^{+5.38}_{-9.12}$ & $0.297^{+0.277}_{-0.158}$~ & $64.61^{+16.67}_{-11.14}$ & $0.260^{+0.285}_{-0.171}$ & - & - \\
    \hline
	ET+CE & $67.70^{+0.33}_{-0.26}$ & $0.308^{+0.009}_{-0.007}$~ & $67.94^{+2.83}_{-3.41}$ & $0.312^{+0.057}_{-0.064}$ & ~~$-1.02^{+0.31}_{-0.25}$ & ~~$0.39^{+1.01}_{-1.55}$~ \\
    \hline
    \hline
    \end{tabular}    
    \caption{Cosmological parameter constraints with $1\sigma$ uncertainty in the $\Lambda$CDM and the $w_0w_a$CDM model from marginalized posteriors averaged over all realizations.}
    \label{tab:results}
\end{table*}
\begin{table*}
    \centering
    \begin{tabular}{c|cccc|ccc}
    \hline
    \hline
     & \multicolumn{4}{c}{($\Xi_0,n$)} & \multicolumn{3}{c}{$\alpha$-basis} \\
    \hline
     & ~$H_0~[{\rm km}$ ${\rm s}^{-1}{\rm Mpc}^{-1}]$ & $\Omega_{m,0}$ & $\Xi_0$ & $n$ & ~$H_0~[{\rm km}$ ${\rm s}^{-1}{\rm Mpc}^{-1}]$ & $\Omega_{m,0}$ &  $c_M$ \\
    \hline
    LVK O5 & $66.38^{+10.35}_{-10.80}$ & - & ~~$1.07^{+0.58}_{-0.47}$ & ~~$0.21^{+6.21}_{-0.21}$~ & $~65.44^{+10.04}_{-11.79}~$ & - & ~~$0.15^{+1.57}_{-1.30}$~ \\
    \hline
    ~LVK post-O5~ & $65.30^{+6.56}_{-7.37}$ & $0.587^{+0.354}_{-0.260}$ & ~~$1.13^{+0.39}_{-0.43}$ & $0.21^{+6.44}_{-0.21}$ & $~64.77^{+6.48}_{-7.54}~$ & - & ~~$0.48^{+0.98}_{-1.12}$ \\
    \hline
	ET+CE & $67.68^{+0.41}_{-0.56}$ & $0.310^{+0.040}_{-0.038}$ & ~~$1.00^{+0.09}_{-0.07}$ & $0.23^{+6.37}_{-0.23}$ & $~67.69^{+0.38}_{-0.51}~$ & $0.309^{+0.043}_{-0.037}$ & ~~$0.01^{+0.11}_{-0.12}$ \\
    \hline
    \hline
    \end{tabular}    
    \caption{Cosmological parameter constraints with $1\sigma$ uncertainty in modified gravity models from marginalized posteriors averaged over all realizations.}
    \label{tab:results_2}
\end{table*}

\section{Conclusion}
\label{sec:conclusion}

In this work, we generalize GW lensing cosmography by associating doubly lensed GWs with strong lensing systems in galaxy surveys. By jointly re-analyzing the two lensed GW images using the SIS lens model, we can measure $d_L^{\rm GW}$ for the unlensed waveform and the impact parameter $y$. Under precise lensed GW localization with the effective detector network, the lens galaxy and the host galaxy of lensed GWs are very likely to be identified from strong lensing catalogues, which could provide redshift of the lens and the source, as well as Einsteins radius in the SIS model from the two image positions. With these observables, we can constrain cosmological parameters and modified GW propagation effects by combining time-delay cosmography and the standard siren approach. We simulate lensed GWs from BBH mergers detected by the LVK O5/post-O5 network and the ET+CE network, and select events within the range and resolvable by the LSST given its design, without using real or mock LSST data.

By simulating 1000 realizations of 5-year observations, we find that the chance to observe lensed GWs by the LVK O5 network that can be associated with LSST is low, with an average detection number of 0.17 over all realizations. But once such events are detected, they are expected to provide moderate constraints on $H_0$ with $14\%$ uncertainty and modified GW propagation effects. With the LVK post-O5 network, the average detection number of lensed GWs associated with LSST increases to 2.3 over all realizations, which slightly enhance the cosmological constraints, but dynamical dark energy remains unconstrained. However, using the ET+CE network, we can detect an average of 80.9 lensed GWs associated with LSST over 100 realizations. They can provide strong constraint on $H_0$ with $0.42\%$ uncertainty and on $\Omega_{m,0}$ with $2.6\%$ uncertainty in the $\Lambda$CDM model. It can also provide moderate constraint on $w_0$ and $w_a$ in the $w_0w_a$CDM model for dynamical dark energy, yielding $w_0=-1.02^{+0.31}_{-0.25}$ and $w_a=0.39^{+1.01}_{-1.55}$. It demonstrates the potential of our method to give an independent measurement of $H_0$ and shed light on the Hubble tension problem using the SIS lens model. Furthermore, it could also jointly constrain modified GW propagation effects with a high precision.

As discussed in literature, a small portion of strongly lensed GWs can produce more than two images. For triply lensed or quadruply lensed GWs, one can reconstruct the lens system with a more comprehensive lens model, such as the singular isothermal ellipsoid (SIE) model, using software like \texttt{lenstronomy} \cite{2015ApJ...813..102B}. With 10 quadruply lensed GWs detected by ET+CE and the images of their lensing systems, one can also constrain $H_0$ with an uncertainty $<1\%$ \cite{Chen:2025xeg}. Although the constraint by a single doubly lensed event is less accurate than a quadruply lensed event due to the difficulty in reconstructing the Fermat potential, the combination of a larger number of doubly lensed events can still provide a competitive constraint. 

However, there exists some limitations in our method as well. First, the localization of doubly lensed GW events is more challenging than quadruply lensed events, so that the identification of the true lens systems with EM images would be more difficult, especially for the LVK network when the GW sky localization is not sufficiently precise. The detectability of lensed events also depends on accurate matching between lensed signals, careful lens model selection \cite{Wright:2023npv}, and accounting for the selection effects of lensed signals (e.g. in \cite{Seo:2026eto}). Second, in reality, reconstruction of the lens model would be less accurate when the EM images have irregular shapes such as arcs, so that a clear image position angle is difficult to obtain. Third, using the SIS model for elliptical lens galaxies would lead to significant biases in cosmography.
Therefore, despite of larger difficulty in lens reconstruction with only two images, a more careful choice is to use a more complex lens model than the SIS model for strong lensing by galaxies even for doubly lensed events, which however would lead to higher degeneracies between lens model parameters and cosmology, and thus weaker constraints on cosmology. Nevertheless, a combination of doubly lensed GWs and quadruply lensed GWs will further enhance the cosmological constraints.

Apart from direct constraints from strongly lensed GWs and their EM counterparts in galaxy catalogues, there are other proposals to constrain cosmology and modified gravity using lensed GWs as well, for example, using lensed GW population and time-delay distribution \cite{Jana:2022shb,Jana:2024uta,Maity:2025apt,Ying:2025eem}, or 
cross-correlation between weak-lensing of GWs and galaxy surveys \cite{Mukherjee:2019wcg,Balaudo:2022znx}. These independent methods could further provide great complements to the cosmological constraints with GW lensing. In conclusion, with the promising future of next-generation GW detectors and large scale galaxy surveys, GW lensing cosmography will be a powerful tool in the era of precision cosmology.

\begin{acknowledgments}
We thank Eungwang Seo for insightful comments during the internal review of our manuscript. A.C. is supported by the China Postdoctoral Science Foundation under Grant No. 2025M773325, and the National Natural Science Foundation of China (NSFC) under Grant No. E414660101 and 12147103. J.Z. is supported by the NSFC under Grants No.~E414660101 and No.~12147103, and the Fundamental Research Funds for the Central Universities under Grants No.~E4EQ6604X2 and No.~E3ER6601A2. This material is
based upon work supported by NSF’s LIGO Laboratory which is a major facility fully funded by the National Science Foundation. We are grateful to the High Performance Computing Center (HPCC) of ICTP-AP for performing the numerical computations in this paper.
\end{acknowledgments}

The data that support the findings of this article are openly available \cite{data}.

\appendix

\section{GW population model}
\label{app:model}

In our work, we simulate mock BBH events with a population model in the space of mass and redshift given by
\begin{equation}
    p_{\rm pop}(m_1^s,m_2^s,z|\Lambda_m,\Lambda_r) \propto p(m_1^s,m_2^s|\Lambda_m)p_{\rm rate}(z|\Lambda_r)\frac{{\rm d}V_c}{{\rm d}z},
\end{equation}
where $m_1^s,m_2^s$ are the source-frame mass. ${\rm d}V_c/{\rm d}z$ is the differential comoving volume. We adopt the Power-law + Double Peak black hole mass distribution model, which is a phenomenological model well fitted by GWTC-4 data \cite{LIGOScientific:2025jau}. It is described by 10 parameters: minimal mass $M_{\rm min}$, maximal mass $M_{\rm max}$, slope of the power-law component of the primary mass distribution $\alpha$, fraction of both Gaussian components in the primary mass distribution, $\lambda_{\rm g}$, fraction of the lower Gaussian component in the two Gaussian components $\lambda_{\rm g,low}$, mean of the lower and higher Gaussian component of the primary mass distribution $\mu_{\rm g,low}$, $\mu_{\rm g,high}$, width of the lower and higher Gaussian component of the primary mass distribution, $\sigma_{\rm g,low}$, $\sigma_{\rm g,high}$, and the range of mass tapering at the lower end of the mass distribution $\delta_m$. The primary mass prior for this population model is given by
\begin{align}
    p(m_1^s|\Lambda_m) & = (1-\lambda_g){\cal B}(m_1^s|M_{\rm min},M_{\rm max},\alpha) \nonumber\\
    & +\lambda_g\lambda_{\rm g,low}{\cal G}(m_1^s|\mu_{\rm g,low},\sigma_{\rm g,low}) \nonumber\\
    & +\lambda_g(1-\lambda_{\rm g,low}){\cal G}(m_1^s|\mu_{\rm g,high},\sigma_{\rm g,high}),
\end{align}
where ${\cal B}$ represents the broken power-law distribution, and ${\cal G}$ is the Gaussian distribution. The secondary mass is then drawn with a power-law mass ratio distribution with a slope of $\beta$.

\begin{figure}
\centering
\begin{subfigure}{0.47\textwidth}
\includegraphics[width=\textwidth]{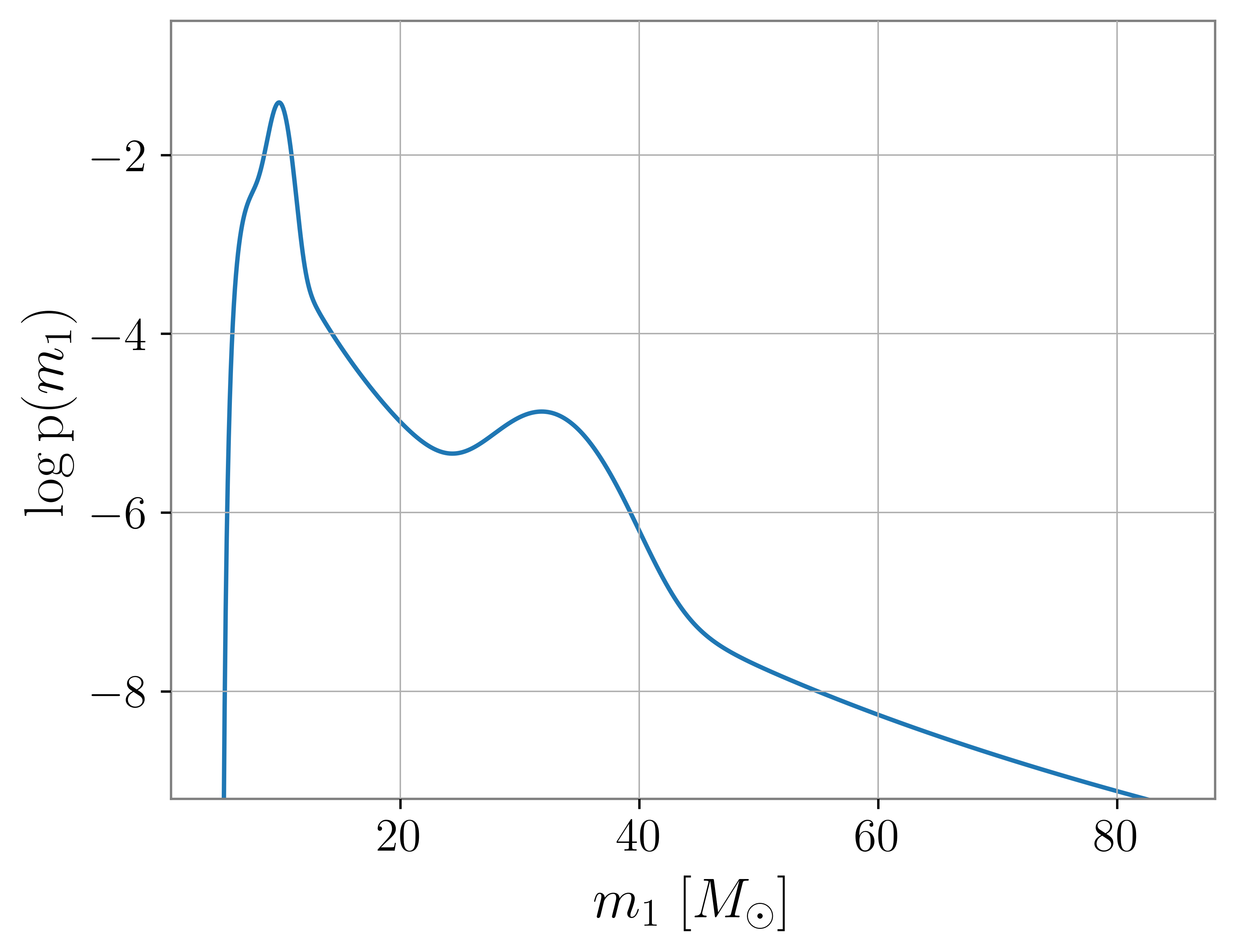}
\end{subfigure}
\begin{subfigure}{0.49\textwidth}
\includegraphics[width=\textwidth]{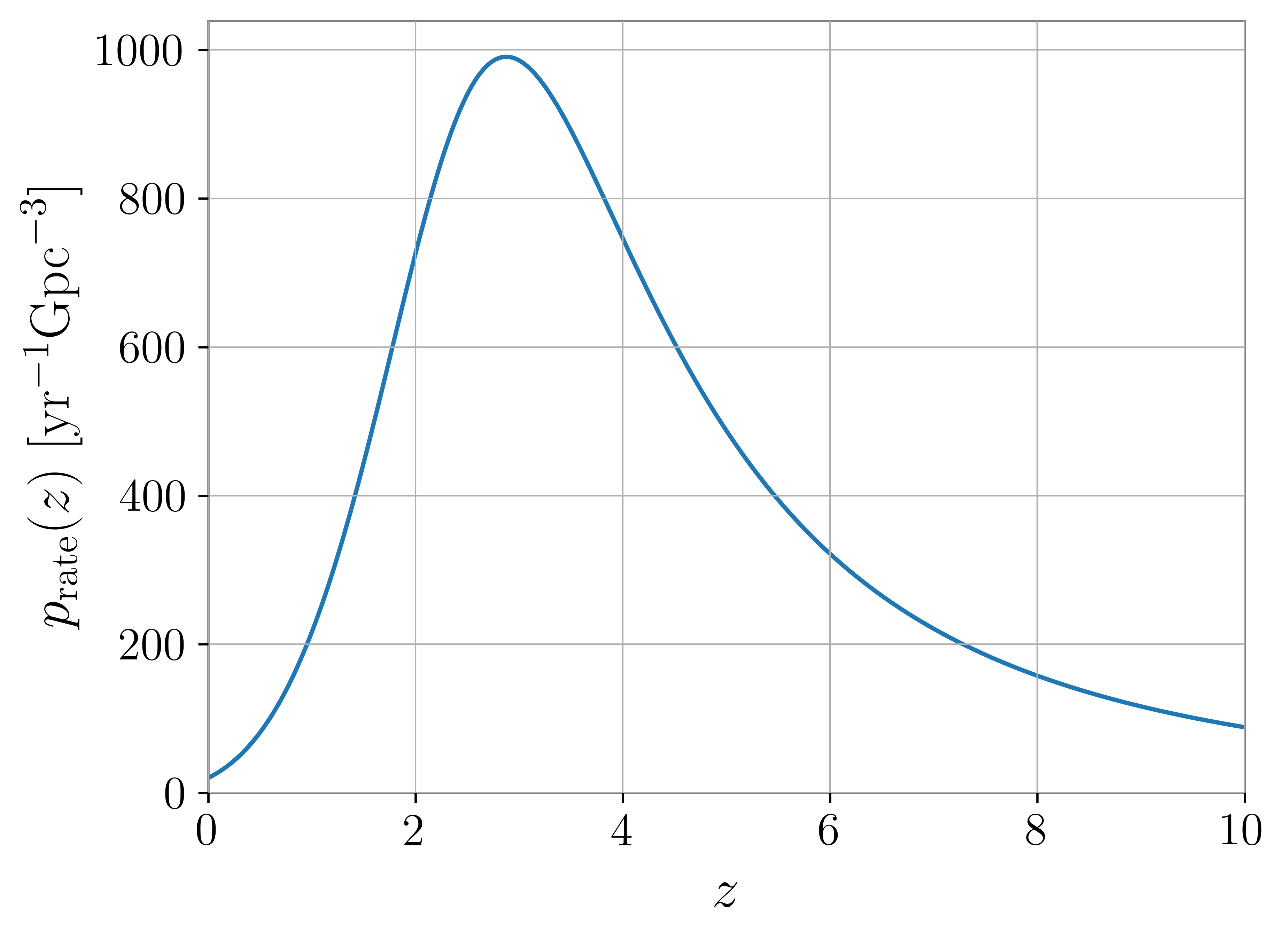}
\end{subfigure}
\caption{Upper panel: Powerlaw + Double Peak black hole mass distribution model. Lower panel: Madau-Dickinson merger-rate redshift evolution model. Both figures are plotted with the parameter values from GWTC-4 cosmology analysis \cite{LIGOScientific:2025jau}.}
\label{fig:Pm1_Pz}
\end{figure}

We also apply the Madau-Dickinson merger rate redshift evolution model inspired from the Madau-Dickinson star formation rate \cite{Madau:2014bja} to our simulation, which is used by the GWTC-4 cosmology paper \cite{LIGOScientific:2025jau}. The merger rate in function of redshift is given by
\begin{equation}
    p_{\rm rate}(z|\Lambda_r) = R_0 [1+(1+z_p)^{-\gamma-k}] \frac{(1+z)^\gamma}{1+\left(\frac{1+z}{1+z_p}\right)^{\gamma+k}},
\end{equation}
where $R_0$ is the merger rate at $z=0$, $\gamma$ and $k$ are the slopes of the powerlaw rate evolution at redshift lower and higher than the turning point $z_p$, respectively.
In our simulation, we adopt the parameter values from GWTC-4 cosmology analysis \cite{LIGOScientific:2025jau}, which are
$M_{\rm min}=4.72M_\odot$, $M_{\rm max}=84.28M_\odot$, $\alpha=2.97$, $\beta=0.83$, $\mu_{\rm g,low}=9.91M_\odot$, $\sigma_{\rm g,low}=0.83M_\odot$, $\lambda_{\rm g,low}=0.86$, $\mu_{\rm g,high}=32.31M_\odot$, $\sigma_{\rm g,high}=4.24M_\odot$, $\lambda_{\rm g,high}=0.28$, $\delta_m=4.71M_\odot$, $\gamma=3.46$, $k=2.90$, and $z_p=2.77$, as listed in Table \ref{tab:GWTC4_param}. For the local merger rate density, we adopt a typical value $R_0=20{\rm yr}^{-1}{\rm Gpc}^{-3}$. The corresponding mass distribution and redshift evolution are plotted in Figure \ref{fig:Pm1_Pz}. 

\renewcommand{\arraystretch}{1.5}
\begin{table*}
    \centering
    \caption{Parameters for the Power-law + Double Peak black hole mass distribution model and the Madau-Dickinson merger rate redshift evolution model used in our mock event simulation.}
    \label{tab:GWTC4_param}
    \begin{tabular}{c|c|c|c|c|c|c|c|c|c|c|c|c|c}
        \hline
        $M_{\rm min}[M_\odot]$ & $M_{\rm max}[M_\odot]$ & $\alpha$ & $\beta$ & $\mu_{\rm g,low}[M_\odot]$ & $\sigma_{\rm g,low}[M_\odot]$ & $\lambda_{\rm g,low}$ & $\mu_{\rm g,high}[M_\odot]$ & $\sigma_{\rm g,high}[M_\odot]$ & $\lambda_{\rm g,high}$ & $\delta_m[M_\odot]$ & $\gamma$ & $k$ & $z_p$ \\
        \hline
        $4.72$ & $84.28$ & $~2.97~$ & $~0.83~$ & $9.91$ & $0.83$ & $0.86$ & $32.31$ & $4.24$ & $0.28$ & $4.71$ & $~3.46~$ & $~2.90~$ & $~2.77$ \\
        \hline
    \end{tabular}
\end{table*}

\section{Optical depth of sources}
\label{app:optical_depth}

\begin{figure}[t]
\centering
\includegraphics[width=0.49\textwidth]{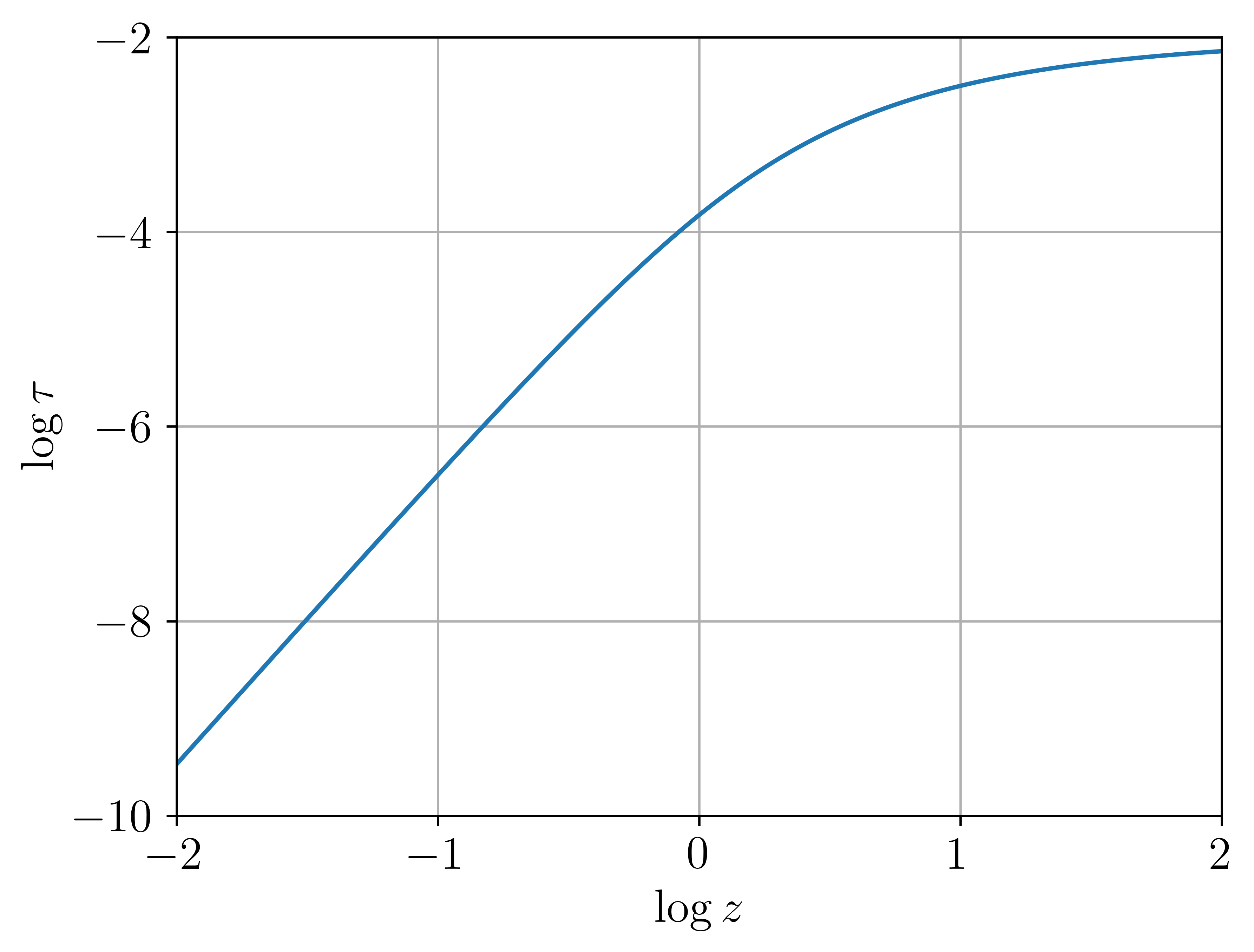}
\caption{Optical depth of background sources computed with equation (\ref{eq:tau_z}) using $n_{*,0}=8.0\times10^{-3}~h^3{\rm Mpc}^{-3}$, $\sigma_{*,0}=144~{\rm km}~{\rm s}^{-1}$, $\alpha=2.49$, and $\beta=2.29$.}
\label{fig:tau_z}
\end{figure}
\begin{figure}
\centering
\includegraphics[width=0.49\textwidth]{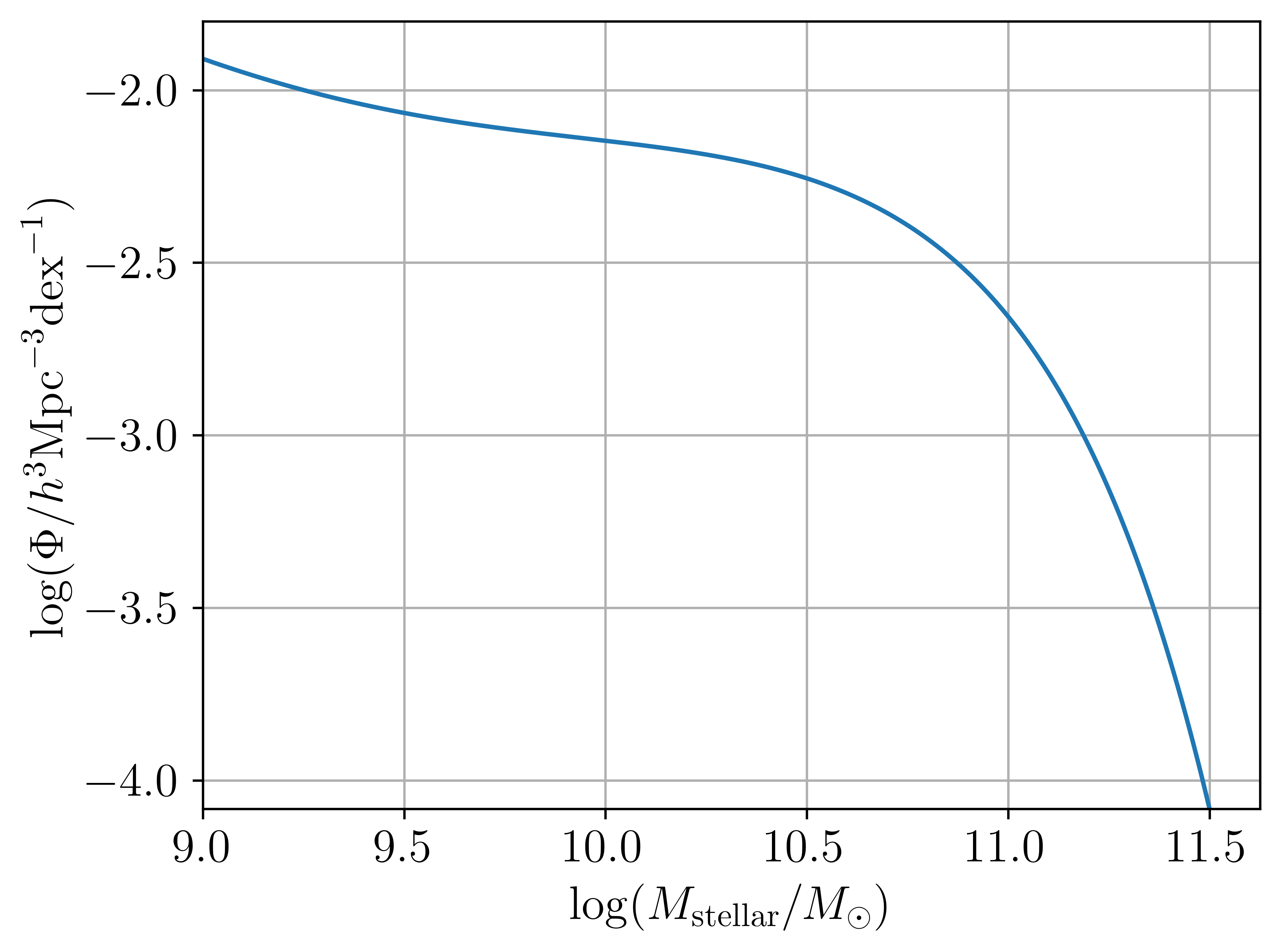}
\caption{The density profile of lens galaxies in stellar masses computed with equation (\ref{eq:double_schechter}).}
\label{fig:phi_M}
\end{figure}
The probability of a background source to be lensed by a foreground galaxy is also known as the optical depth. Its differentiation with respect to the redshift and the velocity dispersion of the lens galaxy is give by \cite{Sereno:2011ty}
\begin{equation}
    \frac{{\rm d}^2\tau}{{\rm d}z_L{\rm d}\sigma} = \frac{{\rm d}n}{{\rm d}\sigma}(z_L, \sigma)s_{\rm cr}(z_L, \sigma) \frac{c{\rm d}t}{{\rm d}z_L}(z_L),
    \label{eq:differential_tau}
\end{equation}
where $s_{\rm cr}$ is the cross section of the signal deflection in the lens system. The lens number density distribution with respect to the galaxy velocity dispersion $\sigma$ can be modelled by a modified Schechter function \cite{SDSS:2003fyb} as
\begin{equation}
    \frac{{\rm d}n}{{\rm d}\sigma} = n_{*}\bigg(\frac{\sigma}{\sigma_*} \bigg)^{\alpha} \exp\bigg[-\bigg(\frac{\sigma}{\sigma_*} \bigg)^{\beta} \bigg] \frac{\beta}{\Gamma[\alpha/\beta]}\frac{1}{\sigma},
\end{equation}
where $\alpha$ is the faint-end slope, $\beta$ is the high-velocity cut-off, and $n_{*}$ and $\sigma_*$ are the characteristic number density and velocity dispersion, which could be parameterized in a form evolving with redshift as $n_*(z)=n_{*,0}(1+z)^{3-\nu_{n^*}}$ and $\sigma_*(z)=\sigma_{*,0}(1+z)^{\nu_{\sigma^*}}$. In our work, we adopt the constant comoving lens number density, so that $\nu_{n^*}=\nu_{\sigma^*}=0$.

The lensing statistic in a finite transient observation needs to account for the missing events due to the time delay of lensed signals \cite{Oguri:2002hv}. The SIS cross section in a transient survey is given by
\begin{equation}
    s_{\rm cr}(z_L, \sigma) = \pi D_L^2\theta_E^2 \int_{y_{\rm min}}^{y_{\rm max}} f(\Delta t(y;\sigma,z_L))y{\rm d}y,
    \label{eq:cross_section}
\end{equation}
where $f(\Delta t)$ is the fraction of lensed events with time delay $\Delta t$ that can be observed. Assuming a uniform distribution of lensed signal arrival time during the continuous transient observation time $T_{\rm obs}$, we have
\begin{equation}
    f(\Delta t) = 1-\frac{\Delta t}{T_{\rm obs}}
\end{equation}
for $\Delta t<T_{\rm obs}$, and $f(\Delta t)=0$ otherwise. Then the cross section in equation (\ref{eq:cross_section}) results in
\begin{equation}
    s_{\rm cr} = \pi D_L^2\theta_E^2 \bigg[(y_{\rm max}^2-y_{\rm min}^2)-\frac{2}{3}\frac{\Delta t_z}{T_{\rm obs}}(y_{\rm max}^3-y_{\rm min}^3) \bigg].
\end{equation}
For the SIS model, we have $y_{\rm min}=0$. By inserting the above equation into equation (\ref{eq:differential_tau}) and integrating it over $\sigma$ in $(0,\infty)$, we arrive at
\begin{equation}
    \tau = \frac{F_*}{30}[D_S(1+z_S)]^3 y_{\rm max}^2 \bigg\{ 1-\frac{1}{7}\frac{\Gamma[(8+\alpha)/\beta]}{\Gamma[(4+\alpha)/\beta]}\frac{\Delta t_*}{T_{\rm obs}} \bigg\},
    \label{eq:tau_z}
\end{equation}
where
\begin{equation}
    F_* = 16\pi^3 n_{*,0} \bigg(\frac{\sigma_{*,0}}{c} \bigg)^4 \frac{\Gamma[(4+\alpha)/\beta]}{\Gamma[\alpha/\beta]};
\end{equation}
\begin{equation}
    \Delta t_* = 32\pi^2 \bigg(\frac{\sigma_{*,0}}{c} \bigg)^4 \frac{D_S}{c}(1+z_S)y_{\rm max},
\end{equation}
Figure \ref{fig:tau_z} shows the optical depth computed with parameters used in Section \ref{sec:lens_select}. In addition, \ref{fig:phi_M} shows the density profile of lens galaxies in stellar masses described by the double Schechter function in equation (\ref{eq:double_schechter}).

\bibliography{reference}

@article{Planck:2018vyg,
    author = "Aghanim, N. and others",
    collaboration = "Planck",
    title = "{Planck 2018 results. VI. Cosmological parameters}",
    eprint = "1807.06209",
    archivePrefix = "arXiv",
    primaryClass = "astro-ph.CO",
    doi = "10.1051/0004-6361/201833910",
    journal = "Astron. Astrophys.",
    volume = "641",
    pages = "A6",
    year = "2020",
    note = "[Erratum: Astron.Astrophys. 652, C4 (2021)]"
}

@article{Riess:2019cxk,
    author = "Riess, Adam G. and Casertano, Stefano and Yuan, Wenlong and Macri, Lucas M. and Scolnic, Dan",
    title = "{Large Magellanic Cloud Cepheid Standards Provide a 1\% Foundation for the Determination of the Hubble Constant and Stronger Evidence for Physics beyond $\Lambda$CDM}",
    eprint = "1903.07603",
    archivePrefix = "arXiv",
    primaryClass = "astro-ph.CO",
    doi = "10.3847/1538-4357/ab1422",
    journal = "Astrophys. J.",
    volume = "876",
    number = "1",
    pages = "85",
    year = "2019"
}

@article{DESI:2025zgx,
    author = "Abdul Karim, M. and others",
    collaboration = "DESI",
    title = "{DESI DR2 results. II. Measurements of baryon acoustic oscillations and cosmological constraints}",
    eprint = "2503.14738",
    archivePrefix = "arXiv",
    primaryClass = "astro-ph.CO",
    reportNumber = "FERMILAB-PUB-25-0169-PPD",
    doi = "10.1103/tr6y-kpc6",
    journal = "Phys. Rev. D",
    volume = "112",
    number = "8",
    pages = "083515",
    year = "2025"
}

@article{DESI:2025wyn,
    author = "Gu, Gan and others",
    collaboration = "DESI",
    title = "{Dynamical Dark Energy in light of the DESI DR2 Baryonic Acoustic Oscillations Measurements}",
    eprint = "2504.06118",
    archivePrefix = "arXiv",
    primaryClass = "astro-ph.CO",
    reportNumber = "FERMILAB-PUB-25-0235-PPD",
    doi = "10.1038/s41550-025-02669-6",
    journal = {Nature Astronomy},
         year = 2025,
        month = sep
}

@article{Abdalla:2022yfr,
    author = "Abdalla, Elcio and others",
    title = "{Cosmology intertwined: A review of the particle physics, astrophysics, and cosmology associated with the cosmological tensions and anomalies}",
    eprint = "2203.06142",
    archivePrefix = "arXiv",
    primaryClass = "astro-ph.CO",
    reportNumber = "FERMILAB-CONF-22-192-SCD",
    doi = "10.1016/j.jheap.2022.04.002",
    journal = "JHEAp",
    volume = "34",
    pages = "49--211",
    year = "2022"
}

@article{Schutz1986,
    author = "Schutz, B.",
    title = "Determining the Hubble constant from gravitational wave observations",
    journal = "Nature",
    year = "1986",
    volume = "323",
    pages = "310-311",
    doi = "10.1038/323310a0",
    url = "https://doi.org/10.1038/323310a0",
}

@article{LIGOScientific:2017adf,
    author = "Abbott, B. P. and others",
    collaboration = "LIGO Scientific, Virgo, 1M2H, Dark Energy Camera GW-E, DES, DLT40, Las Cumbres Observatory, VINROUGE, MASTER",
    title = "{A gravitational-wave standard siren measurement of the Hubble constant}",
    eprint = "1710.05835",
    archivePrefix = "arXiv",
    primaryClass = "astro-ph.CO",
    reportNumber = "LIGO-P1700296, FERMILAB-PUB-17-472-A-AE",
    doi = "10.1038/nature24471",
    journal = "Nature",
    volume = "551",
    number = "7678",
    pages = "85--88",
    year = "2017"
}

@article{Mastrogiovanni:2023emh,
    author = "Mastrogiovanni, Simone and Laghi, Danny and Gray, Rachel and Santoro, Giada Caneva and Ghosh, Archisman and Karathanasis, Christos and Leyde, Konstantin and Steer, Daniele A. and Perries, Stephane and Pierra, Gregoire",
    title = "{Joint population and cosmological properties inference with gravitational waves standard sirens and galaxy surveys}",
    eprint = "2305.10488",
    archivePrefix = "arXiv",
    primaryClass = "astro-ph.CO",
    doi = "10.1103/PhysRevD.108.042002",
    journal = "Phys. Rev. D",
    volume = "108",
    number = "4",
    pages = "042002",
    year = "2023"
}

@article{Gray:2023wgj,
    author = "Gray, Rachel and others",
    title = "{Joint cosmological and gravitational-wave population inference using dark sirens and galaxy catalogues}",
    eprint = "2308.02281",
    archivePrefix = "arXiv",
    primaryClass = "astro-ph.CO",
    doi = "10.1088/1475-7516/2023/12/023",
    journal = "JCAP",
    volume = "12",
    pages = "023",
    year = "2023"
}

@article{Chen:2023wpj,
    author = "Chen, Anson and Gray, Rachel and Baker, Tessa",
    title = "{Testing the nature of gravitational wave propagation using dark sirens and galaxy catalogues}",
    eprint = "2309.03833",
    archivePrefix = "arXiv",
    primaryClass = "gr-qc",
    doi = "10.1088/1475-7516/2024/02/035",
    journal = "JCAP",
    volume = "02",
    pages = "035",
    year = "2024"
}

@article{LIGOScientific:2021aug,
    author = "Abbott, R. and others",
    collaboration = "LIGO Scientific, Virgo, KAGRA",
    title = "{Constraints on the Cosmic Expansion History from GWTC{\textendash}3}",
    eprint = "2111.03604",
    archivePrefix = "arXiv",
    primaryClass = "astro-ph.CO",
    reportNumber = "LIGO-P2100185-v6, LIGO-P2100185-v5",
    doi = "10.3847/1538-4357/ac74bb",
    journal = "Astrophys. J.",
    volume = "949",
    number = "2",
    pages = "76",
    year = "2023"
}

@misc{LIGOScientific:2025slb,
    author = "Abac, A. G. and others",
    collaboration = "LIGO Scientific, VIRGO, KAGRA",
    title = "{GWTC-4.0: Updating the Gravitational-Wave Transient Catalog with Observations from the First Part of the Fourth LIGO-Virgo-KAGRA Observing Run}",
    eprint = "2508.18082",
    archivePrefix = "arXiv",
    primaryClass = "gr-qc",
    reportNumber = "LIGO-P2400386",
    month = "8",
    year = "2025"
}

@article{LIGOScientific:2025jau,
    author = "Abac, A. G. and others",
    collaboration = "LIGO Scientific, VIRGO, KAGRA",
    title = "{GWTC-4.0: Constraints on the Cosmic Expansion Rate and Modified Gravitational-wave Propagation}",
    eprint = "2509.04348",
    archivePrefix = "arXiv",
    primaryClass = "astro-ph.CO",
    reportNumber = "LIGO-P2400152",
    journal = "arXiv:2509.04348",
    year = "2025"
}

@article{Chen:2024gdn,
    author = "Chen, Hsin-Yu and Ezquiaga, Jose Mar{\'\i}a and Gupta, Ish",
    title = "{Cosmography with next-generation gravitational wave detectors}",
    eprint = "2402.03120",
    archivePrefix = "arXiv",
    primaryClass = "gr-qc",
    doi = "10.1088/1361-6382/ad424f",
    journal = "Class. Quant. Grav.",
    volume = "41",
    number = "12",
    pages = "125004",
    year = "2024"
}

@article{Chen:2025qsl,
    author = "Chen, Anson",
    title = "{Measuring the cosmic dipole with golden dark sirens in the era of next-generation ground-based gravitational wave detectors}",
    eprint = "2505.12678",
    archivePrefix = "arXiv",
    primaryClass = "gr-qc",
    doi = "10.1088/1475-7516/2025/07/076",
    journal = "JCAP",
    volume = "07",
    pages = "076",
    year = "2025"
}

@article{Refsdal:1964blz,
    author = "Refsdal, Sjur",
    title = "{On the Possibility of Determining Hubble's Parameter and the Masses of Galaxies from the Gravitational Lens Effect}",
    doi = "10.1093/mnras/128.4.307",
    journal = "Mon. Not. Roy. Astron. Soc.",
    volume = "128",
    number = "4",
    pages = "307--310",
    year = "1964"
}

@inproceedings{TDCOSMO:2025dmr,
    author = "Birrer, Simon and others",
    collaboration = "TDCOSMO",
    title = "{TDCOSMO 2025: Cosmological constraints from strong lensing time delays}",
    booktitle = "{3rd General Meeting of CosmoVerse}: {Addressing observational tensions in cosmology with systematics and fundamental physics}",
    eprint = "2506.03023",
    archivePrefix = "arXiv",
    primaryClass = "astro-ph.CO",
    reportNumber = "FERMILAB-PUB-25-0381-CSAID",
    month = "6",
    year = "2025"
}

@article{Treu:2016ljm,
    author = "Treu, Tommaso and Marshall, Philip J.",
    title = "{Time Delay Cosmography}",
    eprint = "1605.05333",
    archivePrefix = "arXiv",
    primaryClass = "astro-ph.CO",
    doi = "10.1007/s00159-016-0096-8",
    journal = "Astron. Astrophys. Rev.",
    volume = "24",
    number = "1",
    pages = "11",
    year = "2016"
}

@article{Birrer:2020tax,
    author = "Birrer, S. and others",
    title = "{TDCOSMO - IV. Hierarchical time-delay cosmography {\textendash} joint inference of the Hubble constant and galaxy density profiles}",
    eprint = "2007.02941",
    archivePrefix = "arXiv",
    primaryClass = "astro-ph.CO",
    doi = "10.1051/0004-6361/202038861",
    journal = "Astron. Astrophys.",
    volume = "643",
    pages = "A165",
    year = "2020"
}

@article{Birrer:2022chj,
    author = "Birrer, S. and Millon, M. and Sluse, D. and Shajib, A. J. and Courbin, F. and Erickson, S. and Koopmans, L. V. E. and Suyu, S. H. and Treu, T.",
    title = "{Time-Delay Cosmography: Measuring the Hubble Constant and Other Cosmological Parameters with Strong Gravitational Lensing}",
    eprint = "2210.10833",
    archivePrefix = "arXiv",
    primaryClass = "astro-ph.CO",
    doi = "10.1007/s11214-024-01079-w",
    journal = "Space Sci. Rev.",
    volume = "220",
    number = "5",
    pages = "48",
    year = "2024"
}

@article{Huang:2023prq,
    author = "Huang, Shun-Jia and Hu, Yi-Ming and Chen, Xian and Zhang, Jian-dong and Li, En-Kun and Gao, Zucheng and Lin, Xin-Yi",
    title = "{Measuring the Hubble constant using strongly lensed gravitational wave signals}",
    eprint = "2304.10435",
    archivePrefix = "arXiv",
    primaryClass = "astro-ph.CO",
    doi = "10.1088/1475-7516/2023/08/003",
    journal = "JCAP",
    volume = "08",
    pages = "003",
    year = "2023"
}

@article{Suyu:2023jue,
    author = "Suyu, Sherry H. and Goobar, Ariel and Collett, Thomas and More, Anupreeta and Vernardos, Giorgos",
    title = "{Strong Gravitational Lensing and Microlensing of Supernovae}",
    eprint = "2301.07729",
    archivePrefix = "arXiv",
    primaryClass = "astro-ph.CO",
    doi = "10.1007/s11214-024-01044-7",
    journal = "Space Sci. Rev.",
    volume = "220",
    number = "1",
    pages = "13",
    year = "2024"
}

@article{Sereno:2011ty,
    author = "Sereno, M. and Jetzer, Ph. and Sesana, A. and Volonteri, M.",
    title = "{Cosmography with strong lensing of LISA gravitational wave sources}",
    eprint = "1104.1977",
    archivePrefix = "arXiv",
    primaryClass = "astro-ph.CO",
    doi = "10.1111/j.1365-2966.2011.18895.x",
    journal = "Mon. Not. Roy. Astron. Soc.",
    volume = "415",
    pages = "2773",
    year = "2011"
}

@article{Liao:2017ioi,
    author = "Liao, Kai and Fan, Xi-Long and Ding, Xu-Heng and Biesiada, Marek and Zhu, Zong-Hong",
    title = "{Precision cosmology from future lensed gravitational wave and electromagnetic signals}",
    eprint = "1703.04151",
    archivePrefix = "arXiv",
    primaryClass = "astro-ph.CO",
    doi = "10.1038/s41467-017-01152-9",
    journal = "Nature Commun.",
    volume = "8",
    number = "1",
    pages = "1148",
    year = "2017",
    note = "[Erratum: Nature Commun. 8, 2136 (2017)]"
}

@article{Hannuksela:2020xor,
    author = "Hannuksela, Otto A. and Collett, Thomas E. and {\c{C}}al{\i}{\c{s}}kan, Mesut and Li, Tjonnie G. F.",
    title = "{Localizing merging black holes with sub-arcsecond precision using gravitational-wave lensing}",
    eprint = "2004.13811",
    archivePrefix = "arXiv",
    primaryClass = "astro-ph.HE",
    doi = "10.1093/mnras/staa2577",
    journal = "Mon. Not. Roy. Astron. Soc.",
    volume = "498",
    number = "3",
    pages = "3395--3402",
    year = "2020"
}

@article{Wempe:2022zlk,
    author = "Wempe, Ewoud and Koopmans, L{\'e}on V. E. and Wierda, A. Renske A. C. and Hannuksela, Otto Akseli and van den Broeck, Chris",
    title = "{On the detection and precise localization of merging black holes events through strong gravitational lensing}",
    eprint = "2204.08732",
    archivePrefix = "arXiv",
    primaryClass = "astro-ph.HE",
    doi = "10.1093/mnras/stae1023",
    journal = "Mon. Not. Roy. Astron. Soc.",
    volume = "530",
    number = "3",
    pages = "3368--3390",
    year = "2024"
}

@article{Chen:2025xeg,
    author = "Chen, Zhiwei and Yu, Qingjuan and Lu, Youjun and Guo, Xiao",
    title = "{Enhanced Localization of Dark Lensed Gravitational-wave Events Enables Host Galaxy Identification and Precise Cosmological Inference}",
    eprint = "2510.12470",
    archivePrefix = "arXiv",
    primaryClass = "astro-ph.CO",
    doi = "10.3847/2041-8213/ae1226",
    journal = "Astrophys. J. Lett.",
    volume = "993",
    number = "2",
    pages = "L57",
    year = "2025"
}

@article{Li:2018prc,
    author = "Li, Shun-Sheng and Mao, Shude and Zhao, Yuetong and Lu, Youjun",
    title = "{Gravitational lensing of gravitational waves: A statistical perspective}",
    eprint = "1802.05089",
    archivePrefix = "arXiv",
    primaryClass = "astro-ph.CO",
    doi = "10.1093/mnras/sty411",
    journal = "Mon. Not. Roy. Astron. Soc.",
    volume = "476",
    number = "2",
    pages = "2220--2229",
    year = "2018"
}

@article{Branchesi:2023mws,
    author = "Branchesi, Marica and others",
    title = "{Science with the Einstein Telescope: a comparison of different designs}",
    eprint = "2303.15923",
    archivePrefix = "arXiv",
    primaryClass = "gr-qc",
    reportNumber = "ET-0084A-23",
    doi = "10.1088/1475-7516/2023/07/068",
    journal = "J. Cosmol. Astropart. Phys.",
    volume = "07",
    pages = "068",
    year = "2023"
}

@misc{Evans:2021gyd,
    author = "Evans, Matthew and others",
    title = "{A Horizon Study for Cosmic Explorer: Science, Observatories, and Community}",
    eprint = "2109.09882",
    archivePrefix = "arXiv",
    primaryClass = "astro-ph.IM",
    reportNumber = "CE-P2100003-v7, Cosmic Explorer technical report CE-P2100003-v6",
    month = "9",
    year = "2021"
}

@article{Euclid:2025,
   title={Euclid Quick Data Release (Q1). The Strong Lensing Discovery Engine C: Finding lenses with machine learning},
   ISSN={1432-0746},
   url={http://dx.doi.org/10.1051/0004-6361/202554542},
   DOI={10.1051/0004-6361/202554542},
   journal={Astron. Astrophys.},
   publisher={EDP Sciences},
   author={Lines, N.E.P. and others},
   year={2025},
   month=jun }

@article{Shajib:2024yft,
    author = "Shajib{\textdagger}, Anowar J. and Smith, Graham P. and Birrer, Simon and Verma, Aprajita and Arendse, Nikki and Collett, Thomas E. and Daylan, Tansu and Serjeant, Stephen",
    collaboration = "LSST Strong Lensing Science",
    title = "{Strong gravitational lenses from the Vera C. Rubin Observatory}",
    eprint = "2406.08919",
    archivePrefix = "arXiv",
    primaryClass = "astro-ph.CO",
    doi = "10.1098/rsta.2024.0117",
    journal = "Phil. Trans. Roy. Soc. Lond. A",
    volume = "383",
    number = "2295",
    pages = "20240117",
    year = "2025"
}

@BOOK{1987gady.book.....B,
       author = {{Binney}, James and {Tremaine}, Scott},
        title = "{Galactic dynamics}",
         year = 1987,
       adsurl = {https://ui.adsabs.harvard.edu/abs/1987gady.book.....B}
}

@article{Takahashi_2003,
doi = {10.1086/377430},
url = {https://dx.doi.org/10.1086/377430},
year = {2003},
month = {oct},
publisher = {},
volume = {595},
number = {2},
pages = {1039},
author = {Ryuichi Takahashi and Takashi Nakamura},
title = {Wave Effects in the Gravitational Lensing of Gravitational Waves from Chirping Binaries},
journal = {The Astrophysical Journal}
}

@article{Takahashi:2016jom,
    author = "Takahashi, Ryuichi",
    title = "{Arrival time differences between gravitational waves and electromagnetic signals due to gravitational lensing}",
    eprint = "1606.00458",
    archivePrefix = "arXiv",
    primaryClass = "astro-ph.CO",
    doi = "10.3847/1538-4357/835/1/103",
    journal = "Astrophys. J.",
    volume = "835",
    number = "1",
    pages = "103",
    year = "2017"
}

@article{H0LiCOW:2019pvv,
    author = "Wong, Kenneth C. and others",
    collaboration = "H0LiCOW",
    title = "{H0LiCOW {\textendash} XIII. A 2.4 per cent measurement of H0 from lensed quasars: 5.3{\ensuremath{\sigma}} tension between early- and late-Universe probes}",
    eprint = "1907.04869",
    archivePrefix = "arXiv",
    primaryClass = "astro-ph.CO",
    doi = "10.1093/mnras/stz3094",
    journal = "Mon. Not. Roy. Astron. Soc.",
    volume = "498",
    number = "1",
    pages = "1420--1439",
    year = "2020"
}

@article{Belgacem:2018lbp,
    author = "Belgacem, Enis and Dirian, Yves and Foffa, Stefano and Maggiore, Michele",
    title = "{Modified gravitational-wave propagation and standard sirens}",
    eprint = "1805.08731",
    archivePrefix = "arXiv",
    primaryClass = "gr-qc",
    doi = "10.1103/PhysRevD.98.023510",
    journal = "Phys. Rev. D",
    volume = "98",
    number = "2",
    pages = "023510",
    year = "2018"
}

@article{LISACosmologyWorkingGroup:2019mwx,
    author = "Belgacem, Enis and others",
    collaboration = "LISA Cosmology Working Group",
    title = "{Testing modified gravity at cosmological distances with LISA standard sirens}",
    eprint = "1906.01593",
    archivePrefix = "arXiv",
    primaryClass = "astro-ph.CO",
    reportNumber = "LISA CosWG-19-01; IFT-UAM-CSIC-19-79, LISA CosWG-19-01",
    doi = "10.1088/1475-7516/2019/07/024",
    journal = "JCAP",
    volume = "07",
    pages = "024",
    year = "2019"
}

@article{Lagos:2019kds,
    author = "Lagos, Macarena and Fishbach, Maya and Landry, Philippe and Holz, Daniel E.",
    title = "{Standard sirens with a running Planck mass}",
    eprint = "1901.03321",
    archivePrefix = "arXiv",
    primaryClass = "astro-ph.CO",
    doi = "10.1103/PhysRevD.99.083504",
    journal = "Phys. Rev. D",
    volume = "99",
    number = "8",
    pages = "083504",
    year = "2019"
}

@article{Finke:2021aom,
    author = "Finke, Andreas and Foffa, Stefano and Iacovelli, Francesco and Maggiore, Michele and Mancarella, Michele",
    title = "{Cosmology with LIGO/Virgo dark sirens: Hubble parameter and modified gravitational wave propagation}",
    eprint = "2101.12660",
    archivePrefix = "arXiv",
    primaryClass = "astro-ph.CO",
    doi = "10.1088/1475-7516/2021/08/026",
    journal = "JCAP",
    volume = "08",
    pages = "026",
    year = "2021"
}

@article{Finke:2021znb,
    author = "Finke, Andreas and Foffa, Stefano and Iacovelli, Francesco and Maggiore, Michele and Mancarella, Michele",
    title = "{Probing modified gravitational wave propagation with strongly lensed coalescing binaries}",
    eprint = "2107.05046",
    archivePrefix = "arXiv",
    primaryClass = "gr-qc",
    doi = "10.1103/PhysRevD.104.084057",
    journal = "Phys. Rev. D",
    volume = "104",
    number = "8",
    pages = "084057",
    year = "2021"
}

@article{Leyde:2022orh,
    author = "Leyde, Konstantin and Mastrogiovanni, Simone and Steer, Dani{\`e}le A. and Chassande-Mottin, Eric and Karathanasis, Christos",
    title = "{Current and future constraints on cosmology and modified gravitational wave friction from binary black holes}",
    eprint = "2202.00025",
    archivePrefix = "arXiv",
    primaryClass = "gr-qc",
    doi = "10.1088/1475-7516/2022/09/012",
    journal = "JCAP",
    volume = "09",
    pages = "012",
    year = "2022"
}

@ARTICLE{2015ApJ...813..102B,
       author = {{Birrer}, Simon and {Amara}, Adam and {Refregier}, Alexandre},
        title = "{Gravitational Lens Modeling with Basis Sets}",
      journal = {\apj},
         year = 2015,
        month = nov,
       volume = {813},
       number = {2},
          eid = {102},
        pages = {102},
          doi = {10.1088/0004-637X/813/2/102},
archivePrefix = {arXiv},
       eprint = {1504.07629},
 primaryClass = {astro-ph.CO},
       adsurl = {https://ui.adsabs.harvard.edu/abs/2015ApJ...813..102B}
}

@article{Ezquiaga:2020gdt,
    author = "Ezquiaga, Jose Mar{\'\i}a and Holz, Daniel E. and Hu, Wayne and Lagos, Macarena and Wald, Robert M.",
    title = "{Phase effects from strong gravitational lensing of gravitational waves}",
    eprint = "2008.12814",
    archivePrefix = "arXiv",
    primaryClass = "gr-qc",
    doi = "10.1103/PhysRevD.103.064047",
    journal = "Phys. Rev. D",
    volume = "103",
    number = "6",
    pages = "064047",
    year = "2021"
}

@article{Li:2023zdl,
    author = "Li, Alvin K. Y. and Chan, Juno C. L. and Fong, Heather and Chong, Aidan H. Y. and Weinstein, Alan J. and Ezquiaga, Jose M.",
    title = "{TESLA-X: An effective method to search for sub-threshold lensed gravitational waves with a targeted population model}",
    eprint = "2311.06416",
    archivePrefix = "arXiv",
    primaryClass = "gr-qc",
    doi = "10.1093/mnras/staf1259",
    journal = "Mon. Not. Roy. Astron. Soc.",
    volume = "998",
    pages = "1010",
    year = "2025"
}

@article{Li:2019osa,
    author = "Li, Alvin K. Y. and Lo, Rico K. L. and Sachdev, Surabhi and Chan, Juno C. L. and Lin, E. T. and Li, Tjonnie G. F. and Weinstein, Alan J.",
    collaboration = "LIGO Scientific, Virgo",
    title = "{Targeted subthreshold search for strongly lensed gravitational-wave events}",
    eprint = "1904.06020",
    archivePrefix = "arXiv",
    primaryClass = "gr-qc",
    doi = "10.1103/PhysRevD.107.123014",
    journal = "Phys. Rev. D",
    volume = "107",
    number = "12",
    pages = "123014",
    year = "2023"
}

@article{McIsaac:2019use,
    author = "McIsaac, Connor and Keitel, David and Collett, Thomas and Harry, Ian and Mozzon, Simone and Edy, Oliver and Bacon, David",
    title = "{Search for strongly lensed counterpart images of binary black hole mergers in the first two LIGO observing runs}",
    eprint = "1912.05389",
    archivePrefix = "arXiv",
    primaryClass = "gr-qc",
    reportNumber = "LIGO-P1900360",
    doi = "10.1103/PhysRevD.102.084031",
    journal = "Phys. Rev. D",
    volume = "102",
    number = "8",
    pages = "084031",
    year = "2020"
}

@article{LIGOScientific:2023bwz,
    author = "Abbott, R. and others",
    collaboration = "LIGO Scientific, KAGRA, VIRGO",
    title = "{Search for Gravitational-lensing Signatures in the Full Third Observing Run of the LIGO{\textendash}Virgo Network}",
    eprint = "2304.08393",
    archivePrefix = "arXiv",
    primaryClass = "gr-qc",
    reportNumber = "LIGO-P2200031",
    doi = "10.3847/1538-4357/ad3e83",
    journal = "Astrophys. J.",
    volume = "970",
    number = "2",
    pages = "191",
    year = "2024"
}

@article{Narola:2023viz,
    author = {Narola, Harsh and Janquart, Justin and Haegel, Le{\"\i}la and Haris, K. and Hannuksela, Otto A. and Van Den Broeck, Chris},
    title = "{How well can modified gravitational wave propagation be constrained with strong lensing?}",
    eprint = "2308.01709",
    archivePrefix = "arXiv",
    primaryClass = "gr-qc",
    doi = "10.1103/PhysRevD.109.084064",
    journal = "Phys. Rev. D",
    volume = "109",
    number = "8",
    pages = "084064",
    year = "2024"
}

@book{Dodelson:2003ft,
    author = "Dodelson, Scott",
    title = "{Modern Cosmology}",
    isbn = "978-0-12-219141-1",
    publisher = "Academic Press",
    address = "Amsterdam",
    year = "2003"
}

@article{Chevallier:2000qy,
    author = "Chevallier, Michel and Polarski, David",
    title = "{Accelerating universes with scaling dark matter}",
    eprint = "gr-qc/0009008",
    archivePrefix = "arXiv",
    doi = "10.1142/S0218271801000822",
    journal = "Int. J. Mod. Phys. D",
    volume = "10",
    pages = "213--224",
    year = "2001"
}

@article{Linder:2002et,
    author = "Linder, Eric V.",
    title = "{Exploring the expansion history of the universe}",
    eprint = "astro-ph/0208512",
    archivePrefix = "arXiv",
    doi = "10.1103/PhysRevLett.90.091301",
    journal = "Phys. Rev. Lett.",
    volume = "90",
    pages = "091301",
    year = "2003"
}

@article{deRham:2018red,
    author = "de Rham, Claudia and Melville, Scott",
    title = "{Gravitational Rainbows: LIGO and Dark Energy at its Cutoff}",
    eprint = "1806.09417",
    archivePrefix = "arXiv",
    primaryClass = "hep-th",
    reportNumber = "Imperial/TP/2018/CdR/02",
    doi = "10.1103/PhysRevLett.121.221101",
    journal = "Phys. Rev. Lett.",
    volume = "121",
    number = "22",
    pages = "221101",
    year = "2018"
}

@article{LISACosmologyWorkingGroup:2022wjo,
    author = "Baker, Tessa and others",
    collaboration = "LISA Cosmology Working Group",
    title = "{Measuring the propagation speed of gravitational waves with LISA}",
    eprint = "2203.00566",
    archivePrefix = "arXiv",
    primaryClass = "gr-qc",
    reportNumber = "LISA CosWG-22-02",
    doi = "10.1088/1475-7516/2022/08/031",
    journal = "JCAP",
    volume = "08",
    number = "08",
    pages = "031",
    year = "2022"
}

@article{Baker:2022eiz,
    author = "Baker, Tessa and Barausse, Enrico and Chen, Anson and de Rham, Claudia and Pieroni, Mauro and Tasinato, Gianmassimo",
    title = "{Testing gravitational wave propagation with multiband detections}",
    eprint = "2209.14398",
    archivePrefix = "arXiv",
    primaryClass = "gr-qc",
    doi = "10.1088/1475-7516/2023/03/044",
    journal = "JCAP",
    volume = "03",
    pages = "044",
    year = "2023"
}

@article{LIGOScientific:2017vwq,
    author = "Abbott, B. P. and others",
    collaboration = "LIGO Scientific, Virgo",
    title = "{GW170817: Observation of Gravitational Waves from a Binary Neutron Star Inspiral}",
    eprint = "1710.05832",
    archivePrefix = "arXiv",
    primaryClass = "gr-qc",
    reportNumber = "LIGO-P170817",
    doi = "10.1103/PhysRevLett.119.161101",
    journal = "Phys. Rev. Lett.",
    volume = "119",
    number = "16",
    pages = "161101",
    year = "2017"
}

@article{LIGOScientific:2017zic,
    author = "Abbott, B. P. and others",
    collaboration = "LIGO Scientific, Virgo, Fermi-GBM, INTEGRAL",
    title = "{Gravitational Waves and Gamma-rays from a Binary Neutron Star Merger: GW170817 and GRB 170817A}",
    eprint = "1710.05834",
    archivePrefix = "arXiv",
    primaryClass = "astro-ph.HE",
    reportNumber = "LIGO-P1700308",
    doi = "10.3847/2041-8213/aa920c",
    journal = "Astrophys. J. Lett.",
    volume = "848",
    number = "2",
    pages = "L13",
    year = "2017"
}

@article{LIGOScientific:2017ync,
    author = "Abbott, B. P. and others",
    collaboration = "LIGO Scientific, Virgo, Fermi GBM, INTEGRAL, IceCube, AstroSat Cadmium Zinc Telluride Imager Team, IPN, Insight-Hxmt, ANTARES, Swift, AGILE Team, 1M2H Team, Dark Energy Camera GW-EM, DES, DLT40, GRAWITA, Fermi-LAT, ATCA, ASKAP, Las Cumbres Observatory Group, OzGrav, DWF (Deeper Wider Faster Program), AST3, CAASTRO, VINROUGE, MASTER, J-GEM, GROWTH, JAGWAR, CaltechNRAO, TTU-NRAO, NuSTAR, Pan-STARRS, MAXI Team, TZAC Consortium, KU, Nordic Optical Telescope, ePESSTO, GROND, Texas Tech University, SALT Group, TOROS, BOOTES, MWA, CALET, IKI-GW Follow-up, H.E.S.S., LOFAR, LWA, HAWC, Pierre Auger, ALMA, Euro VLBI Team, Pi of Sky, Chandra Team at McGill University, DFN, ATLAS Telescopes, High Time Resolution Universe Survey, RIMAS, RATIR, SKA South Africa/MeerKAT",
    title = "{Multi-messenger Observations of a Binary Neutron Star Merger}",
    eprint = "1710.05833",
    archivePrefix = "arXiv",
    primaryClass = "astro-ph.HE",
    reportNumber = "LIGO-P1700294, VIR-0802A-17, FERMILAB-PUB-17-478-A-AE-CD",
    doi = "10.3847/2041-8213/aa91c9",
    journal = "Astrophys. J. Lett.",
    volume = "848",
    number = "2",
    pages = "L12",
    year = "2017"
}

@article{Nishizawa:2017nef,
    author = "Nishizawa, Atsushi",
    title = "{Generalized framework for testing gravity with gravitational-wave propagation. I. Formulation}",
    eprint = "1710.04825",
    archivePrefix = "arXiv",
    primaryClass = "gr-qc",
    doi = "10.1103/PhysRevD.97.104037",
    journal = "Phys. Rev. D",
    volume = "97",
    number = "10",
    pages = "104037",
    year = "2018"
}

@article{Arai:2017hxj,
    author = "Arai, Shun and Nishizawa, Atsushi",
    title = "{Generalized framework for testing gravity with gravitational-wave propagation. II. Constraints on Horndeski theory}",
    eprint = "1711.03776",
    archivePrefix = "arXiv",
    primaryClass = "gr-qc",
    doi = "10.1103/PhysRevD.97.104038",
    journal = "Phys. Rev. D",
    volume = "97",
    number = "10",
    pages = "104038",
    year = "2018"
}

@article{Amendola:2017ovw,
    author = "Amendola, Luca and Sawicki, Ignacy and Kunz, Martin and Saltas, Ippocratis D.",
    title = "{Direct detection of gravitational waves can measure the time variation of the Planck mass}",
    eprint = "1712.08623",
    archivePrefix = "arXiv",
    primaryClass = "astro-ph.CO",
    doi = "10.1088/1475-7516/2018/08/030",
    journal = "JCAP",
    volume = "08",
    pages = "030",
    year = "2018"
}

@article{Romano:2023ozy,
    author = "Romano, Antonio Enea and Sakellariadou, Mairi",
    title = "{Mirage of Luminal Modified Gravitational-Wave Propagation}",
    eprint = "2302.05413",
    archivePrefix = "arXiv",
    primaryClass = "gr-qc",
    reportNumber = "CERN-TH-2023-017, KCL-PH-TH-2023-08",
    doi = "10.1103/PhysRevLett.130.231401",
    journal = "Phys. Rev. Lett.",
    volume = "130",
    number = "23",
    pages = "231401",
    year = "2023"
}

@article{Belgacem:2017ihm,
    author = "Belgacem, Enis and Dirian, Yves and Foffa, Stefano and Maggiore, Michele",
    title = "{Gravitational-wave luminosity distance in modified gravity theories}",
    eprint = "1712.08108",
    archivePrefix = "arXiv",
    primaryClass = "astro-ph.CO",
    doi = "10.1103/PhysRevD.97.104066",
    journal = "Phys. Rev. D",
    volume = "97",
    number = "10",
    pages = "104066",
    year = "2018"
}

@article{Belgacem:2017cqo,
    author = "Belgacem, Enis and Dirian, Yves and Foffa, Stefano and Maggiore, Michele",
    title = "{Nonlocal gravity. Conceptual aspects and cosmological predictions}",
    eprint = "1712.07066",
    archivePrefix = "arXiv",
    primaryClass = "hep-th",
    doi = "10.1088/1475-7516/2018/03/002",
    journal = "JCAP",
    volume = "03",
    pages = "002",
    year = "2018"
}

@article{Bellini:2014fua,
    author = "Bellini, Emilio and Sawicki, Ignacy",
    title = "{Maximal freedom at minimum cost: linear large-scale structure in general modifications of gravity}",
    eprint = "1404.3713",
    archivePrefix = "arXiv",
    primaryClass = "astro-ph.CO",
    doi = "10.1088/1475-7516/2014/07/050",
    journal = "JCAP",
    volume = "07",
    pages = "050",
    year = "2014"
}

@article{Linder:2015rcz,
    author = {Linder, Eric V. and Seng{\"o}r, Gizem and Watson, Scott},
    title = "{Is the Effective Field Theory of Dark Energy Effective?}",
    eprint = "1512.06180",
    archivePrefix = "arXiv",
    primaryClass = "astro-ph.CO",
    doi = "10.1088/1475-7516/2016/05/053",
    journal = "JCAP",
    volume = "05",
    pages = "053",
    year = "2016"
}

@article{Noller:2018wyv,
    author = "Noller, Johannes and Nicola, Andrina",
    title = "{Cosmological parameter constraints for Horndeski scalar-tensor gravity}",
    eprint = "1811.12928",
    archivePrefix = "arXiv",
    primaryClass = "astro-ph.CO",
    doi = "10.1103/PhysRevD.99.103502",
    journal = "Phys. Rev. D",
    volume = "99",
    number = "10",
    pages = "103502",
    year = "2019"
}

@article{Bellini:2015xja,
    author = "Bellini, Emilio and Cuesta, Antonio J. and Jimenez, Raul and Verde, Licia",
    title = "{Constraints on deviations from {\ensuremath{\Lambda}}CDM within Horndeski gravity}",
    eprint = "1509.07816",
    archivePrefix = "arXiv",
    primaryClass = "astro-ph.CO",
    doi = "10.1088/1475-7516/2016/06/E01",
    journal = "JCAP",
    volume = "02",
    pages = "053",
    year = "2016",
    note = "[Erratum: JCAP 06, E01 (2016)]"
}

@article{Ezquiaga:2021ayr,
    author = "Ezquiaga, Jose Mar{\'\i}a",
    title = "{Hearing gravity from the cosmos: GWTC-2 probes general relativity at cosmological scales}",
    eprint = "2104.05139",
    archivePrefix = "arXiv",
    primaryClass = "astro-ph.CO",
    doi = "10.1016/j.physletb.2021.136665",
    journal = "Phys. Lett. B",
    volume = "822",
    pages = "136665",
    year = "2021"
}

@article{Ishak:2024jhs,
    author = "Ishak, M. and others",
    title = "{Modified gravity constraints from the full shape modeling of clustering measurements from DESI 2024}",
    eprint = "2411.12026",
    archivePrefix = "arXiv",
    primaryClass = "astro-ph.CO",
    reportNumber = "FERMILAB-PUB-24-0848-PPD",
    doi = "10.1088/1475-7516/2025/09/053",
    journal = "JCAP",
    volume = "09",
    pages = "053",
    year = "2025"
}

@article{Karathanasis:2022hrb,
    author = "Karathanasis, Christos and Revenu, Benoit and Mukherjee, Suvodip and Stachurski, Federico",
    title = "{GWSim: Python package for creating mock GW samples for different astrophysical populations and cosmological models of binary black holes}",
    eprint = "2210.05724",
    archivePrefix = "arXiv",
    primaryClass = "astro-ph.CO",
    doi = "10.1051/0004-6361/202245216",
    journal = "Astron. Astrophys.",
    volume = "677",
    pages = "A124",
    year = "2023",
    note = "[Erratum: Astron.Astrophys. 682, C1 (2024)]"
}

@article{Maoz:1993ix,
    author = "Maoz, Dan and Rix, Hans-Walter",
    title = "{Early type galaxies, dark halos, and gravitational lensing statistics}",
    reportNumber = "IASSNS-AST-93-1",
    doi = "10.1086/173248",
    journal = "Astrophys. J.",
    volume = "416",
    pages = "425",
    year = "1993"
}

@article{Kochanek:1995ap,
    author = "Kochanek, Christopher S.",
    title = "{Is there a cosmological constant?}",
    eprint = "astro-ph/9510077",
    archivePrefix = "arXiv",
    reportNumber = "CFA-4258",
    doi = "10.1086/177538",
    journal = "Astrophys. J.",
    volume = "466",
    pages = "638",
    year = "1996"
}

@article{Mitchell:2004gw,
    author = "Mitchell, Jonathan L. and Keeton, Charles R. and Frieman, Joshua A. and Sheth, Ravi K.",
    title = "{Robust cosmological constraints from gravitational lens statistics}",
    eprint = "astro-ph/0401138",
    archivePrefix = "arXiv",
    reportNumber = "FERMILAB-PUB-04-189-A",
    doi = "10.1086/427910",
    journal = "Astrophys. J.",
    volume = "622",
    pages = "81--98",
    year = "2005"
}

@article{Madau:2014bja,
    author = "Madau, Piero and Dickinson, Mark",
    title = "{Cosmic Star Formation History}",
    eprint = "1403.0007",
    archivePrefix = "arXiv",
    primaryClass = "astro-ph.CO",
    doi = "10.1146/annurev-astro-081811-125615",
    journal = "Ann. Rev. Astron. Astrophys.",
    volume = "52",
    pages = "415--486",
    year = "2014"
}

@ARTICLE{2025MNRAS.537..779F,
       author = {{Ferrami}, G. and {Wyithe}, J. Stuart B.},
        title = "{Velocity dispersion function evolution from strong lensing statistics}",
      journal = "Mon. Not. Roy. Astron. Soc.",
         year = 2025,
        month = feb,
       volume = {537},
       number = {2},
        pages = {779-787},
          doi = {10.1093/mnras/staf089},
archivePrefix = {arXiv},
       eprint = {2410.20892},
 primaryClass = {astro-ph.GA},
       adsurl = {https://ui.adsabs.harvard.edu/abs/2025MNRAS.537..779F}
}

@article{Geng:2021tiz,
    author = "Geng, Shuaibo and Cao, Shuo and Liu, Yuting and Liu, Tonghua and Biesiada, Marek and Lian, Yujie",
    title = "{The velocity dispersion function of early-type galaxies and its redshift evolution: the newest results from lens redshift test}",
    eprint = "2102.12140",
    archivePrefix = "arXiv",
    primaryClass = "astro-ph.GA",
    doi = "10.1093/mnras/stab519",
    journal = "Mon. Not. Roy. Astron. Soc.",
    volume = "503",
    number = "1",
    pages = "1319--1326",
    year = "2021"
}

@article{Choi:2006qg,
    author = "Choi, Yun-Young and Park, Changbom and Vogeley, Michael S.",
    title = "{Internal and Collective Properties of Galaxies in the Sloan Digital Sky Survey}",
    eprint = "astro-ph/0611607",
    archivePrefix = "arXiv",
    doi = "10.1086/511060",
    journal = "Astrophys. J.",
    volume = "658",
    pages = "884--897",
    year = "2007"
}

@article{SDSS:2003fyb,
    author = "Sheth, Ravi K. and others",
    collaboration = "SDSS",
    title = "{The Velocity Dispersion Function of Early-Type Galaxies}",
    eprint = "astro-ph/0303092",
    archivePrefix = "arXiv",
    reportNumber = "FERMILAB-PUB-03-051-A",
    doi = "10.1086/376794",
    journal = "Astrophys. J.",
    volume = "594",
    pages = "225--231",
    year = "2003"
}

@article{Jana:2022shb,
    author = "Jana, Souvik and Kapadia, Shasvath J. and Venumadhav, Tejaswi and Ajith, Parameswaran",
    title = "{Cosmography Using Strongly Lensed Gravitational Waves from Binary Black Holes}",
    eprint = "2211.12212",
    archivePrefix = "arXiv",
    primaryClass = "astro-ph.CO",
    reportNumber = "LIGO-P2200298",
    doi = "10.1103/PhysRevLett.130.261401",
    journal = "Phys. Rev. Lett.",
    volume = "130",
    number = "26",
    pages = "261401",
    year = "2023"
}

@article{Jana:2024uta,
    author = "Jana, Souvik and Kapadia, Shasvath J. and Venumadhav, Tejaswi and More, Surhud and Ajith, Parameswaran",
    title = "{Strong-lensing cosmography using third-generation gravitational-wave detectors}",
    eprint = "2405.17805",
    archivePrefix = "arXiv",
    primaryClass = "gr-qc",
    doi = "10.1088/1361-6382/ad8d2e",
    journal = "Class. Quant. Grav.",
    volume = "41",
    number = "24",
    pages = "245010",
    year = "2024"
}

@misc{Maity:2025apt,
    author = "Maity, Koustav N. and Jana, Souvik and Venumadhav, Tejaswi and Barsode, Ankur and Ajith, Parameswaran",
    title = "{Strong lensing cosmography using binary-black-hole mergers: Prospects for the near future}",
    eprint = "2512.15168",
    archivePrefix = "arXiv",
    primaryClass = "gr-qc",
    month = "12",
    year = "2025"
}

@misc{Ying:2025eem,
    author = "Ying, Xinguang and Yang, Tao",
    title = "{Precision Joint Constraints on Cosmology and Gravity Using Strongly Lensed Gravitational Wave Populations}",
    eprint = "2505.09507",
    archivePrefix = "arXiv",
    primaryClass = "gr-qc",
    month = "5",
    year = "2025"
}

@article{Oguri:2002hv,
    author = "Oguri, Masamune and Suto, Yasushi and Turner, Edwin L.",
    title = "{Gravitational lensing magnification and time delay statistics for distant supernovae}",
    eprint = "astro-ph/0210107",
    archivePrefix = "arXiv",
    reportNumber = "RESCEU-13-02, UTAP-420",
    doi = "10.1086/345431",
    journal = "Astrophys. J.",
    volume = "583",
    pages = "584--593",
    year = "2003"
}

@ARTICLE{2016MNRAS.459.2150W,
       author = {{Weigel}, Anna K. and {Schawinski}, Kevin and {Bruderer}, Claudio},
        title = "{Stellar mass functions: methods, systematics and results for the local Universe}",
      journal = {Mon. Not. Roy. Astron. Soc.},
         year = 2016,
        month = jun,
       volume = {459},
       number = {2},
        pages = {2150-2187},
          doi = {10.1093/mnras/stw756},
archivePrefix = {arXiv},
       eprint = {1604.00008},
 primaryClass = {astro-ph.GA},
       adsurl = {https://ui.adsabs.harvard.edu/abs/2016MNRAS.459.2150W}
}

@misc{LSST_overview,
author = {Gilmore, Keenan and Kahn, Steven and Nordby, Martin and Burke, David and OConnor, Paul and Oliver, John and Radeka, Veljko and Schalk, Terry and Schindler, Rafe},
year = {2006},
month = {01},
pages = {},
title = {The LSST Camera Overview},
url = {https://www.slac.stanford.edu/pubs/slacpubs/12250/slac-pub-12291.pdf}
}

@ARTICLE{2009arXiv0912.0201L,
       author = {{LSST Science Collaboration} and {Abell}, Paul A. and others},
        title = "{LSST Science Book, Version 2.0}",
      journal = {arXiv e-prints},
         year = 2009,
        month = dec,
          eid = {arXiv:0912.0201},
        pages = {arXiv:0912.0201},
          doi = {10.48550/arXiv.0912.0201},
archivePrefix = {arXiv},
       eprint = {0912.0201},
 primaryClass = {astro-ph.IM},
       adsurl = {https://ui.adsabs.harvard.edu/abs/2009arXiv0912.0201L}
}

@misc{LSSTDarkEnergyScience:2012kar,
    author = "Abate, Alexandra and others",
    collaboration = "LSST Dark Energy Science",
    title = "{Large Synoptic Survey Telescope: Dark Energy Science Collaboration}",
    eprint = "1211.0310",
    archivePrefix = "arXiv",
    primaryClass = "astro-ph.CO",
    reportNumber = "FERMILAB-FN-0952-A-T",
    doi = "10.2172/1156445",
    month = "11",
    year = "2012"
}

@article{Pratten:2020ceb,
    author = "Pratten, Geraint and others",
    title = "{Computationally efficient models for the dominant and subdominant harmonic modes of precessing binary black holes}",
    eprint = "2004.06503",
    archivePrefix = "arXiv",
    primaryClass = "gr-qc",
    doi = "10.1103/PhysRevD.103.104056",
    journal = "Phys. Rev. D",
    volume = "103",
    number = "10",
    pages = "104056",
    year = "2021"
}

@article{Williams:2021qyt,
   author = "Williams, Michael J. and Veitch, John and Messenger, Chris",
   title = "{Nested sampling with normalizing flows for gravitational-wave inference}",
   eprint = "2102.11056",
   archivePrefix = "arXiv",
   primaryClass = "gr-qc",
   doi = "10.1103/PhysRevD.103.103006",
   journal = "Phys. Rev. D",
   volume = "103",
   number = "10",
   pages = "103006",
   year = "2021"
}

@article{Williams:2023ppp,
   author = "Williams, Michael J. and Veitch, John and Messenger, Chris",
   title = "{Importance nested sampling with normalising flows}",
   eprint = "2302.08526",
   archivePrefix = "arXiv",
   primaryClass = "astro-ph.IM",
   reportNumber = "LIGO-P2200283",
   doi = "10.1088/2632-2153/acd5aa",
   journal = "Mach. Learn. Sci. Tech.",
   volume = "4",
   number = "3",
   pages = "035011",
   year = "2023"
}

@article{bilby_paper,
    author = "Ashton, Gregory and others",
    title = "{BILBY: A user-friendly Bayesian inference library for gravitational-wave astronomy}",
    eprint = "1811.02042",
    archivePrefix = "arXiv",
    primaryClass = "astro-ph.IM",
    doi = "10.3847/1538-4365/ab06fc",
    journal = "Astrophys. J. Suppl.",
    volume = "241",
    number = "2",
    pages = "27",
    year = "2019"
}

@article{Foreman-Mackey:2012any,
    author = "Foreman-Mackey, Daniel and Hogg, David W. and Lang, Dustin and Goodman, Jonathan",
    title = "{emcee: The MCMC Hammer}",
    eprint = "1202.3665",
    archivePrefix = "arXiv",
    primaryClass = "astro-ph.IM",
    doi = "10.1086/670067",
    journal = "Publ. Astron. Soc. Pac.",
    volume = "125",
    pages = "306--312",
    year = "2013"
}

@article{Balaudo:2022znx,
    author = "Balaudo, Anna and Garoffolo, Alice and Martinelli, Matteo and Mukherjee, Suvodip and Silvestri, Alessandra",
    title = "{Prospects of testing late-time cosmology with weak lensing of gravitational waves and galaxy surveys}",
    eprint = "2210.06398",
    archivePrefix = "arXiv",
    primaryClass = "astro-ph.CO",
    doi = "10.1088/1475-7516/2023/06/050",
    journal = "JCAP",
    volume = "06",
    pages = "050",
    year = "2023"
}

@misc{LIGOScientific:2025cwb,
    author = "Abac, A. G. and others",
    collaboration = "LIGO Scientific, VIRGO, KAGRA",
    title = "{GWTC-4.0: Searches for Gravitational-Wave Lensing Signatures}",
    eprint = "2512.16347",
    archivePrefix = "arXiv",
    primaryClass = "gr-qc",
    reportNumber = "LIGO-P2500419",
    month = "12",
    year = "2025"
}

@article{KAGRA:2013rdx,
    author = "Abbott, B. P. and others",
    collaboration = "KAGRA, LIGO Scientific, Virgo",
    title = "{Prospects for observing and localizing gravitational-wave transients with Advanced LIGO, Advanced Virgo and KAGRA}",
    eprint = "1304.0670",
    archivePrefix = "arXiv",
    primaryClass = "gr-qc",
    reportNumber = "LIGO-P1200087, VIR-0288A-12, JGW-P1808427",
    doi = "10.1007/s41114-020-00026-9",
    journal = "Living Rev. Rel.",
    volume = "19",
    pages = "1",
    year = "2016"
}

@article{Euclid:2024yrr,
    author = "Mellier, Y. and others",
    collaboration = "Euclid",
    title = "{Euclid. I. Overview of the Euclid mission}",
    eprint = "2405.13491",
    archivePrefix = "arXiv",
    primaryClass = "astro-ph.CO",
    doi = "10.1051/0004-6361/202450810",
    journal = "Astron. Astrophys.",
    volume = "697",
    pages = "A1",
    year = "2025"
}

@article{Wright:2023npv,
    author = "Wright, Mick and Janquart, Justin and Hendry, Martin",
    title = "{Determination of Lens Mass Density Profile from Strongly Lensed Gravitational-wave Signals}",
    eprint = "2307.07852",
    archivePrefix = "arXiv",
    primaryClass = "astro-ph.HE",
    doi = "10.3847/1538-4357/ad0891",
    journal = "Astrophys. J.",
    volume = "959",
    number = "2",
    pages = "70",
    year = "2023"
}

@article{Mukherjee:2019wcg,
    author = "Mukherjee, Suvodip and Wandelt, Benjamin D. and Silk, Joseph",
    title = "{Probing the theory of gravity with gravitational lensing of gravitational waves and galaxy surveys}",
    eprint = "1908.08951",
    archivePrefix = "arXiv",
    primaryClass = "astro-ph.CO",
    doi = "10.1093/mnras/staa827",
    journal = "Mon. Not. Roy. Astron. Soc.",
    volume = "494",
    number = "2",
    pages = "1956--1970",
    year = "2020"
}

@article{Mukherjee:2020mha,
    author = "Mukherjee, Suvodip and Wandelt, Benjamin D. and Silk, Joseph",
    title = "{Testing the general theory of relativity using gravitational wave propagation from dark standard sirens}",
    eprint = "2012.15316",
    archivePrefix = "arXiv",
    primaryClass = "astro-ph.CO",
    doi = "10.1093/mnras/stab001",
    journal = "Mon. Not. Roy. Astron. Soc.",
    volume = "502",
    number = "1",
    pages = "1136--1144",
    year = "2021"
}

@article{Garoffolo:2019mna,
    author = "Garoffolo, Alice and Tasinato, Gianmassimo and Carbone, Carmelita and Bertacca, Daniele and Matarrese, Sabino",
    title = "{Gravitational waves and geometrical optics in scalar-tensor theories}",
    eprint = "1912.08093",
    archivePrefix = "arXiv",
    primaryClass = "gr-qc",
    doi = "10.1088/1475-7516/2020/11/040",
    journal = "JCAP",
    volume = "11",
    pages = "040",
    year = "2020"
}

@article{Tasinato:2021wol,
    author = "Tasinato, Gianmassimo and Garoffolo, Alice and Bertacca, Daniele and Matarrese, Sabino",
    title = "{Gravitational-wave cosmological distances in scalar-tensor theories of gravity}",
    eprint = "2103.00155",
    archivePrefix = "arXiv",
    primaryClass = "gr-qc",
    doi = "10.1088/1475-7516/2021/06/050",
    journal = "JCAP",
    volume = "06",
    pages = "050",
    year = "2021"
}

@article{Jin:2025dvf,
    author = "Jin, Shang-Jie and Song, Ji-Yu and Sun, Tian-Yang and Xiao, Si-Ren and Wang, He and Wang, Ling-Feng and Zhang, Jing-Fei and Zhang, Xin",
    title = "{Gravitational wave standard sirens: A brief review of cosmological parameter estimation}",
    eprint = "2507.12965",
    archivePrefix = "arXiv",
    primaryClass = "astro-ph.CO",
    doi = "10.1007/s11433-025-2829-9",
    journal = "Sci. China Phys. Mech. Astron.",
    volume = "69",
    number = "2",
    pages = "220401",
    year = "2026"
}

@article{Zhu:2023rrx,
    author = "Zhu, Tao and Zhao, Wen and Yan, Jian-Ming and Wang, Yuan-Zhu and Gong, Cheng and Wang, Anzhong",
    title = "{Constraints on parity and Lorentz violations in gravity from GWTC-3 through a parametrization of modified gravitational wave propagations}",
    eprint = "2304.09025",
    archivePrefix = "arXiv",
    primaryClass = "gr-qc",
    doi = "10.1103/PhysRevD.110.064044",
    journal = "Phys. Rev. D",
    volume = "110",
    number = "6",
    pages = "064044",
    year = "2024"
}

@article{Zhang:2024rel,
    author = "Zhang, Bo-Yang and Zhu, Tao and Zhang, Jing-Fei and Zhang, Xin",
    title = "{Forecasts for constraining Lorentz-violating damping of gravitational waves from compact binary inspirals}",
    eprint = "2402.08240",
    archivePrefix = "arXiv",
    primaryClass = "gr-qc",
    doi = "10.1103/PhysRevD.109.104022",
    journal = "Phys. Rev. D",
    volume = "109",
    number = "10",
    pages = "104022",
    year = "2024"
}

@article{Lin:2024pkr,
    author = "Lin, Chunbo and Zhu, Tao and Niu, Rui and Zhao, Wen",
    title = "{Constraining the modified friction in gravitational wave propagation with precessing black hole binaries}",
    eprint = "2404.11245",
    archivePrefix = "arXiv",
    primaryClass = "gr-qc",
    doi = "10.1103/s3xh-dt13",
    journal = "Phys. Rev. D",
    volume = "112",
    number = "2",
    pages = "024010",
    year = "2025"
}

@article{Poon:2024zxn,
    author = "Poon, Jason S. C. and Rinaldi, Stefano and Janquart, Justin and Narola, Harsh and Hannuksela, Otto A.",
    title = "{Galaxy lens reconstruction based on strongly lensed gravitational waves: similarity transformation degeneracy and mass-sheet degeneracy}",
    eprint = "2406.06463",
    archivePrefix = "arXiv",
    primaryClass = "astro-ph.HE",
    doi = "10.1093/mnras/stae2660",
    journal = "Mon. Not. Roy. Astron. Soc.",
    volume = "536",
    number = "3",
    pages = "2212--2233",
    year = "2024"
}

@misc{Seo:2026eto,
    author = "Seo, Eungwang and Kim, Kyungmin and Li, Zhuotao and Janquart, Justin and Gray, Rachel and Hendry, Martin",
    title = "{The impact of strong lensing on Hubble constant measurements with gravitational-wave dark sirens}",
    eprint = "2603.01321",
    archivePrefix = "arXiv",
    primaryClass = "astro-ph.CO",
    reportNumber = "LIGO-P2600059",
    month = "3",
    year = "2026"
}

@article{Janquart:2021qov,
    author = "Janquart, Justin and Hannuksela, Otto A. and K., Haris and Van Den Broeck, Chris",
    title = "{A fast and precise methodology to search for and analyse strongly lensed gravitational-wave events}",
    eprint = "2105.04536",
    archivePrefix = "arXiv",
    primaryClass = "gr-qc",
    doi = "10.1093/mnras/stab1991",
    journal = "Mon. Not. Roy. Astron. Soc.",
    volume = "506",
    number = "4",
    pages = "5430--5438",
    year = "2021"
}

@article{Uronen:2024bth,
    author = "Uronen, Laura Elina and Li, Tian and Janquart, Justin and Phurailatpam, Hemanta and Poon, Jason S. C. and Wempe, Ewoud and Koopmans, L{\'e}on V. E. and Hannuksela, Otto A.",
    title = "{Finding black holes: an unconventional multi-messenger}",
    eprint = "2406.14257",
    archivePrefix = "arXiv",
    primaryClass = "astro-ph.HE",
    doi = "10.1098/rsta.2024.0152",
    journal = "Phil. Trans. Roy. Soc. Lond. A",
    volume = "383",
    number = "2294",
    pages = "20240152",
    year = "2025"
}

@article{PhysRevD.46.5236,
  title = {Detection, measurement, and gravitational radiation},
  author = {Finn, Lee S.},
  journal = {Phys. Rev. D},
  volume = {46},
  issue = {12},
  pages = {5236--5249},
  numpages = {0},
  year = {1992},
  month = {Dec},
  publisher = {American Physical Society},
  doi = {10.1103/PhysRevD.46.5236},
  url = {https://link.aps.org/doi/10.1103/PhysRevD.46.5236}
}

@article{PhysRevD.49.2658,
  title = {Gravitational waves from merging compact binaries: How accurately can one extract the binary's parameters from the inspiral waveform?},
  author = {Cutler, Curt and Flanagan, \'Eanna E.},
  journal = {Phys. Rev. D},
  volume = {49},
  issue = {6},
  pages = {2658--2697},
  numpages = {0},
  year = {1994},
  month = {Mar},
  publisher = {American Physical Society},
  doi = {10.1103/PhysRevD.49.2658},
  url = {https://link.aps.org/doi/10.1103/PhysRevD.49.2658}
}

@article{Mancarella:2021ecn,
    author = "Mancarella, Michele and Genoud-Prachex, Edwin and Maggiore, Michele",
    title = "{Cosmology and modified gravitational wave propagation from binary black hole population models}",
    eprint = "2112.05728",
    archivePrefix = "arXiv",
    primaryClass = "gr-qc",
    doi = "10.1103/PhysRevD.105.064030",
    journal = "Phys. Rev. D",
    volume = "105",
    number = "6",
    pages = "064030",
    year = "2022"
}

@misc{data,
  howpublished = {\url{https://github.com/AnChenPhys/GW-lensing-cosmo}} 
}
\bibliographystyle{apsrev4-2}

\end{document}